\documentclass[10pt,conference]{IEEEtran} 
\IEEEoverridecommandlockouts

\usepackage[style=ieee,uniquelist=false,backend=biber,maxcitenames=1,mincitenames=1,minbibnames=1,maxbibnames=4]{biblatex} 
\usepackage{xcolor}

\usepackage{tikz}

\usepackage{nicefrac}

\usepackage{xfrac}

\definecolor{Gray}{gray}{0.3}
\tikzstyle{mybox} = [draw=black, very thick, rectangle, rounded corners, inner ysep=5pt, inner xsep=5pt, fill=gray!20]
\newcommand{\xyz}[2]{
    \smallskip
    \noindent
    \begin{tikzpicture}
        \node [mybox] (box){
            \centering
            \begin{minipage}{.97\columnwidth}
                \fontsize{8.8}{10}\selectfont
                \textbf{RQ #1}. #2
            \end{minipage}
        };
    \end{tikzpicture}
}
\newcommand{\dispbox}[1]{
    \smallskip
    \noindent
    \begin{tikzpicture}
        \node [mybox] (box){
            \centering
            \begin{minipage}{.97\columnwidth}
                \fontsize{8.8}{10}\selectfont
                #1
            \end{minipage}
        };
    \end{tikzpicture}
    \vspace*{-1\baselineskip}
}

\usepackage{amsmath,amssymb}

\usepackage{amsfonts}

\usepackage{graphicx}

\usepackage{textcomp}

\usepackage{wrapfig}

\usepackage{dsfont}

\usepackage{hyperref}

\usepackage[capitalise]{cleveref}
\crefname{figure}{Figure}{Figures}
\crefname{appendixfigure}{Appendix figure}{Appendix figures}

\usepackage{xspace}

\usepackage{paralist}

\usepackage{multirow}

\usepackage{subcaption}

\usepackage{silence}
\usepackage{courier}

\usepackage{mathtools}

\usepackage{nccmath}

\usepackage{listings}
\usepackage{flafter}

\usepackage{pdflscape}

\usepackage[figuresleft]{rotating}

\usepackage[obeyDraft,obeyFinal]{todonotes}

\usepackage{tabularray}

\usepackage{enumitem}

\usepackage{soul}

\hypersetup{final}
\setuptodonotes{inline}
\UseTblrLibrary{booktabs}
\graphicspath{{./figures/}}
\AtBeginDocument{
    \providecommand\BibTeX{{
                Bib\TeX}}}

\newcommand{\sref}[1]{\S~\ref{#1}}

\newcommand{\citet}[1]{\textcite{#1}}
\newcommand{\gptturbo}{\textsc{Gpt-3.5}\xspace}

\newcommand{\ie}{\emph{i.e.},\xspace}
\newcommand{\eg}{\emph{e.g.},\xspace}
\newcommand{\vs}{\emph{vs.}\xspace}
\newcommand{\etc}{\emph{etc.}\xspace}
\newcommand{\viz}{\emph{viz.}\xspace}
\def\BibTeX{{\rm B\kern-.05em{\sc i\kern-.025em b}\kern-.08em
    T\kern-.1667em\lower.7ex\hbox{E}\kern-.125emX}}
\begin{document}
\newcommand{\AverageEceNoScale}{0.32\xspace}
\newcommand{\AverageEceScaled}{0.03\xspace}
\newcommand{\AverageEceScaledMinSkill}{0.05\xspace}
\newcommand{\AverageEceNoScaleMinSkill}{0.38\xspace}
\newcommand{\GptNNTfNormalizationDeltaRawSs}{-0.34\xspace}
\newcommand{\GptNNTfNormalizationDeltaRawEce}{0.19\xspace}
\newcommand{\AverageBestTechnique}{Ask T/F\xspace}
\newcommand{\AverageBestTechniqueSkill}{-1.60\xspace}
\newcommand{\AverageRawSkill}{-7.12\xspace}
\newcommand{\GptNNTtpRawSkill}{-1.07\xspace}
\newcommand{\GptNNTtpRawEce}{0.41\xspace}
\newcommand{\GptNNTtpScaledSkill}{0.05\xspace}
\newcommand{\GptNNTtpScaledEce}{0.05\xspace}
\newcommand{\GptNNTtpMaxDataset}{HumanEval\xspace}
\newcommand{\GptNNTtpBestSs}{0.15\xspace}
\newcommand{\FracHighAccIntro}{52\%\xspace}

\newcommand{\MeanSkillScoreDiffScaled}{0.00\xspace}
\newcommand{\MinSkillScoreDiffScaled}{-0.04\xspace}
\newcommand{\MaxSkillScoreDiffScaled}{0.02\xspace}
\newcommand{\MeanEceDiffScaled}{-0.00\xspace}
\newcommand{\MinEceDiffScaled}{-0.03\xspace}
\newcommand{\MaxEceDiffScaled}{0.03\xspace}
\newcommand{\MeanSkillScoreDiffNonscaled}{-0.11\xspace}
\newcommand{\MinSkillScoreDiffNonscaled}{-0.39\xspace}
\newcommand{\MaxSkillScoreDiffNonscaled}{-0.00\xspace}
\newcommand{\MeanEceDiffNonscaled}{0.02\xspace}
\newcommand{\MinEceDiffNonscaled}{0.01\xspace}
\newcommand{\MaxEceDiffNonscaled}{0.05\xspace}
\newcommand{\GptMeanFracOfRowsFifty}{0.04\xspace}
\newcommand{\GptMaxFracOfRowsFifty}{0.08\xspace}
\newcommand{\GptMinFracOfRowsFifty}{0.02\xspace}
\newcommand{\CodexMeanFracOfRowsFifty}{0.14\xspace}
\newcommand{\CodexMaxFracOfRowsFifty}{0.17\xspace}
\newcommand{\CodexMinFracOfRowsFifty}{0.09\xspace}
\newcommand{\CodegenMeanFracOfRowsFifty}{0.10\xspace}
\newcommand{\CodegenMaxFracOfRowsFifty}{0.12\xspace}
\newcommand{\CodegenMinFracOfRowsFifty}{0.09\xspace}

\title{Calibration and Correctness of\\ Language Models for Code}

\makeatletter 
\newcommand{\linebreakand}{%
  \end{@IEEEauthorhalign}
  \hfill\mbox{}\par
  \mbox{}\hfill\begin{@IEEEauthorhalign}
}
\makeatother 

\author{
\IEEEauthorblockN{Claudio Spiess$^*$}
\IEEEauthorblockA{UC Davis\\
USA\\
cvspiess@ucdavis.edu}

\and

\IEEEauthorblockN{David Gros$^*$}
\IEEEauthorblockA{UC Davis\\
USA\\
dgros@ucdavis.edu}

\and

\IEEEauthorblockN{Kunal Suresh Pai}
\IEEEauthorblockA{UC Davis\\
USA\\
kunpai@ucdavis.edu}

\and

\IEEEauthorblockN{Michael Pradel}
\IEEEauthorblockA{Univ. of Stuttgart\\
Germany\\
michael@binaervarianz.de}

\and

\IEEEauthorblockN{Md Rafiqul Islam Rabin}
\IEEEauthorblockA{Univ. of Houston\\
USA\\
mrabin@central.uh.edu}

\and

\linebreakand

\IEEEauthorblockN{Amin Alipour}
\IEEEauthorblockA{Univ. of Houston\\
USA\\
maalipou@central.uh.edu}

\and

\IEEEauthorblockN{Susmit Jha}
\IEEEauthorblockA{SRI\\
USA\\
susmit.jha@sri.com}

\and

\IEEEauthorblockN{Prem Devanbu}
\IEEEauthorblockA{UC Davis\\
USA\\
ptdevanbu@ucdavis.edu}

\and

\IEEEauthorblockN{Toufique Ahmed}
\IEEEauthorblockA{UC Davis\\
USA\\
tfahmed@ucdavis.edu}
}

\maketitle
\def\thefootnote{*}\footnotetext{Equal contribution. Order determined by random coin flip.}\def\thefootnote{\arabic{footnote}}

\begin{abstract}
Machine learning models are widely used, but can also often be wrong. Users would benefit from a reliable indication of whether a given output from a given model should be trusted, so a rational decision can be made whether to use the output or not. For example, outputs can be associated with a \emph{confidence measure}; if this confidence measure is strongly associated with \emph{likelihood of correctness}, then the model is said to be \emph{well-calibrated}. 

A well-calibrated confidence measure can serve as a basis for rational, graduated decision-making on how much review and care is needed when using generated code.
\emph{Calibration} has so far been studied in mostly non-generative (\eg classification) settings, especially in software engineering. However, generated code can quite often be wrong: Given generated code, developers must decide whether to use directly, use after varying intensity of careful review, or discard model-generated code. Thus, calibration is vital in generative settings.

We make several contributions. We develop a framework for evaluating the calibration of code-generating models. We consider several tasks, correctness criteria, datasets, and approaches, and find that, by and large, generative code models we test are \textbf{\textit{\underline{not}}} well-calibrated out of the box. We then show how calibration can be improved using standard methods, such as Platt scaling. Since Platt scaling relies on the prior availability of correctness data, we evaluate the applicability and generalizability of Platt scaling in software engineering, discuss settings where it has good potential for practical use, and settings where it does not. Our contributions will lead to better-calibrated decision-making in the current use of code generated by language models, and offers a framework for future research to further improve calibration methods for generative models in software engineering.
\end{abstract}

\begin{IEEEkeywords}
LLMs, Calibration, Confidence Measure
\end{IEEEkeywords}

\section{Introduction}
Generative large language models (LLMs) are now widely-used for code completion in IDEs.
However, LLMs make mistakes:
they can generate known buggy code~\cite{jesseLargeLanguageModels2023b},
or code with risky vulnerabilities~\cite{asareGitHubCopilotBad2023b,schusterYouAutocompleteMe2021b}.
Despite these risks,
LLM ``copilots"~\cite{loTrustworthySynergisticArtificial2023a} are growing in popularity---thus there is a growing concern that bad LLM-generated code could be integrated
into widely-used software.
Given that LLMs might generate
buggy code,
how should a developer decide
whether generated code is correct or not?

One possibility is to use the \emph{confidence}, or probability assigned to the generated code by the LLM itself.
Consider a developer who asks \gptturbo to complete some unfinished code. For example, given the prefix\\
{\scriptsize\tt def clear(self, tag: Optional[Hashable] = None) -> None:}\\
the model generates the completion\\
{\scriptsize\tt self.jobs[:] = [job for job in self.jobs if job.tag != tag]}\\
with an average per-token confidence of {\small\tt 91\%}, suggesting high confidence (based on its training) that the code is a likely completion for the given prefix. However, this code is known to be buggy! In fact, when we test thousands of line completions, in cases where the average probability was greater than {\small\tt 90\%}, only {\small\tt \FracHighAccIntro} actually passed test cases. One can also find reverse examples, where the LLM has very little confidence, but the generated code is actually correct.

We make two observations. First, since LLMs can make mistakes, \emph{users would benefit from a reliable indication of confidence that the generated code is actually correct}. Second, \emph{such indications of confidence, when well-aligned with actual correctness, can support well-justified software quality control measures, and thus improve the reliability of the AI4SE ecosystem}.
When an LLM-offered code suggestion is accompanied by a numerical ``confidence'' signal \eg a probability measure, then this signal \emph{should be} well-aligned
with the likelihood that the code
is actually correct. Such a measure
is said to be well\emph{-calibrated}.

Calibration has been studied in other settings \eg classically in weather prediction and recently
for software-related \emph{classification} tasks~\cite{zhouCalibrationPretrainedCode2024}.
In this paper, we
study the calibration of \emph{generative}\footnote{We note that the notion
	of \emph{correctness} for generative tasks is quite different than for classification tasks, where the output is a label, rather than a sequence of tokens.} large language models, when used in practical software engineering
settings, such as line-level code completion, function synthesis, and code-repair.

A well-calibrated confidence measure would support rational risk-management in the development process, and help quality-improvement processes.
For example, a development team might reasonably adopt a policy that: a) generated code associated with high confidence
could be reviewed lightly and quickly accepted; b) suggestions with a medium confidence value should be reviewed more carefully before acceptance; and c) suggestions with a low confidence value should be simply rejected or the prompt should be adjusted.

Despite its importance and widespread use of LLMs in software engineering, the correctness and calibration of code-generating models currently is not well understood.
In particular, it is currently unknown whether confidence measures provided by the LLMs themselves align well with actual code correctness.

This paper does an empirical study of the
calibration of code-generating models using several prior techniques and explores approaches for improving calibration further.
To this end, we describe an evaluation framework for the calibration of code-generating language models.
We instantiate the framework for different tasks, \eg code synthesis, code completion, and program repair, using different correctness criteria (exact-match w.r.t.\ a reference solution and correctness-modulo-testing), and by applying the framework to different models.
Based on this framework, we evaluate  well-established techniques for estimating how a model's confidence align with actual correctness; then, based on our findings, we present improvements over existing techniques.

Our work yields several findings:
\begin{itemize}
	\item
	      The alignment of
	      \emph{confidence measures} provided
	      by LLMs with standard
	      notions of \emph{code correctness} is poor,
	      when evaluated on realistic
	      datasets across different
	      tasks, including completion,
	      synthesis, and automated program repair. We observe generally high $\mathit{ECE}$ (Expected Calibration Error, described in \autoref{sec:calibration_measures}) across all settings, ranging from $0.09$ to $0.73$, suggesting intrinsic LLM confidences are poor predictors of code correctness.
	\item
	      We evaluate several \emph{reflective} approaches
	      to improve this alignment,
	      and also \emph{confidence rescaling} using known correctness labels. While rescaling  generally  improves calibration,
	      reflective methods are rather inconsistent, working better in some settings than others.
	\item
	      Finally, we focus on the most   widely-used SE task for LLMs,
	      \viz code completion,  and use the instructable \gptturbo model,
	      and few-shotting, in a reflective setting, and
	      show that calibration improves substantially
	      from the skill score\footnote{Brier Skill score, which we explain below in~\sref{sec:calibration_measures}.} of $0$ to a much higher level of $0.15$.

\end{itemize}
Our work considers the important problem of providing developers with a reliable indication of whether generated code is correct, and (especially
for the widely-used task of code completion) offers an approach, using BM25-aided~\cite{bm25fewshot} few-shotting, that has potential practical value.

\subsection{Research Agenda}

Code LLMs are perhaps mostly widely-used for \emph{code suggestion/completion}; other tasks include \emph{code synthesis} and \emph{program repair}.

\xyz{1}{How well is the confidence of language models in their output aligned with the
	empirical correctness of the output, specifically for common generative tasks, \viz function synthesis, line-level code completion, and program repair?}
We evaluate the output for two different notions of correctness, \viz, exact-match with known correct code, and second, passing all given tests.

In general, the levels of alignment between the
\emph{intrinsic}
confidence (\emph{viz.,} directly provided by the LLM) and correctness is poor. This indicates need for better approaches to calibration. We then explore
several engineering responses to the problem, listed in the following research questions.
First,  we consider the standard approach of confidence \emph{rescaling}, using Platt scaling.

\xyz{2}{Can alignment between LLM confidence in generated code, and its correctness, be improved by confidence rescaling?}
While confidence rescaling can help remedy over- and under-confidence, it does require some data to determine the parameters of the scaling function. We also analyze and discuss some
considerations in obtaining this data, specifically for code-generation tasks.

Next, we investigate the possibility that the model is able to better calibrate
upon \emph{reflection}. We ask the model (using a separate reflective prompt) to consider its own generated code and judge its confidence in the quality.

\xyz{3}{Is confidence obtained by reflection better aligned with correctness?}
Finally we investigate few-shotting, to see whether it helps calibration for the widely used task of code completion.

\xyz{4}{Can we use few-shot techniques to achieve better calibrated confidence for code completion, using an instruction-tuned model with in-context learning?}
\section{Background}

\subsection{Calibration}
This concept
originates in prediction problems like
weather forecasting. Consider a weather model predicting a 70\% confidence (probability) of rain the next day.
If we ran this model for a while, and observed rain in 70\% of the days where
a forecast with 70\% confidence was made, then we call it a well-calibrated model.
A well-calibrated
model's confidence
in a given output, is quite close to the empirical relative frequency (likelihood) with which the output is actually correct.

With well-calibrated rain-forecasting,
a user has options for a \emph{rational response}:
at 20\% confidence of rain, one might take a hat; at higher confidences, one might take an umbrella; if even higher, one might take the bus rather than walk, \etc
From an earlier work by~\textcite{jiangHowCanWe2021a}:
given a model $M$, an input $X$ and true, expected correct output $Y$,
a model output $M(X) = \hat{Y}$, (note that we won't always have $Y = \hat{Y}$) and a output probability
$P_M (\hat{Y} \,|\, X)$ provided by the model, a perfectly calibrated model satisfies the following condition

\useshortskip
\begin{equation}
	P(\hat{Y} = Y \,|\, P_M (\hat{Y} \,|\, X) = \colorbox{pink}{${p}$}) = \colorbox{yellow}{p}, \quad \forall \, {\mathrm p} \in [0,1].
	\label{eqn:calib}
\end{equation}

In other words, if we have perfectly-calibrated \underline{confidence} $\colorbox{pink}{\footnotesize ${p}$}$ (the model's calculated probability of its prediction that the output is $\hat{Y}$), then this value equals the \emph{empirical fraction} $\colorbox{yellow}{\footnotesize p}$
of the cases where the actual output $Y$ \underline{correctly} matches the prediction $\hat{Y}$.
Usually these probabilities don't perfectly match; there are various measures of the deviation, including \emph{Brier Score}~\cite{brierVerificationForecastsExpressed1950a} and \emph{$ECE$} (Expected
Calibration Error)~\cite{naeiniObtainingWellCalibrated2015a}

\subsection{Why Calibration Matters for Code}
Even powerful
LLMs can make mistakes~\cite{jesseLargeLanguageModels2023b}, 
potentially leading users to accept incorrect code. 
A well-calibrated confidence signal could help developers manage~\cite{loTrustworthySynergisticArtificial2023a, grosAISafetySubproblems2023b} this risk. Consider the confidence
$\colorbox{pink}{\footnotesize ${p}$}$
associated with generated code. A well-calibrated high-value of
$\colorbox{pink}{\footnotesize ${p}$}$
would indicate a high empirical probabilty
$\colorbox{yellow}{\footnotesize p}$
that the code is correct, and so it could be simply accepted;
a low value would indicate
higher risk that the code
is incorrect, and so should be rejected.
A poorly calibrated model may lead to either unnecessary rejection of likely correct code, or ill-advised acceptance of likely incorrect code.
We note that good calibration allows more nuanced, effective quality-control (Q-C) decisions, beyond simple binary decisions \eg carefully review each token of generated code,
perhaps by several people, \vs just use it.
Such a well-calibrated
quality-control process has been used in medicine \eg for elder-care~\cite{niermanOutcomePredictionModel2001a}, and for decision-making
in cancer-care~\cite{schwarzGUESSProjectingMachine2019a}.
Given the cost \& consequences
of properly addressing software quality, and
the potential benefits of LLM-generated code, a well-calibrated confidence signal is highly desirable.

\section{Research Methodology}

We consider three generative tasks \ie function synthesis, line-level code completion, and program repair, where generative LLMs are directly applicable and widely-used (\eg completion in Copilot). In this section, we will discuss the tasks, datasets, models, and methodology of our approach.

\subsection{Code Correctness}
\label{sec:correctness}
When evaluating calibration, we need a notion of correctness. For (non-generative) models outputting labels, classes, True/False, \etc correctness is simply an exact-match with ground-truth correct label. Exact-match \emph{could} also be viewed as a notion of correctness for code, \eg with defect repair, where there is a known, incorrect ``buggy'' version, and a known ``fixed'' version.  Generated code is correct only if it matches exactly the fixed version, given an appropriate prompt. However, this approach is  \emph{\bf overly strict}; the generated code might match exactly, but still pass all tests.
Other notions of correctness exist: code-review, formal verification, \etc We use test cases provided with the code as our preferred indication of correctness, as tests are widely used, and are easily automated.

While test-passing correctness offers the advantage of admitting different semantically identical forms, test cases maybe insufficient or incorrect\footnote{\eg~\citet{liuYourCodeGenerated2023a} report gaps in the test sets of HumanEval.}; tests can also be ``flaky''~\cite{luoEmpiricalAnalysisFlaky2014a}: the same test, on the same code, might pass, or fail.

\subsection{Confidence Measures}
\label{sec:confmeasure}
We usually calculate a confidence measure (or probability) $p$, associated with generated output code $C$.
We consider two categories of measures: \emph{intrinsic probability}, which is calculated by the generative LLM \emph{per se},
and \emph{reflective probability}, obtained by \emph{re-}invoking the model, instructing
it to estimate its confidence in the correctness of the code just generated (see~\cref{app:selfask} for prompts).
Our measures include:
\begin{description}[topsep=10pt,itemsep=5pt,partopsep=4pt,parsep=5pt,leftmargin=0cm,style=unboxed]
	\item[Average Token Probability] \emph{(Intrinsic, $p_{avg}$)} For an output sequence $\mathrm{T}$ of tokens $\tau_i,i = 1 \ldots n$, we collect the associated model probabilities $p(\tau_i)$, and then compute the mean $p_{\text{avg}}(\mathrm{T}) = \frac{1}{n} \displaystyle\sum_{i=1}^n p(\tau_i)$.

	\item[Generated Sequence Probability] \emph{(Intrinsic, $p_{tot}$)} The full generated sequence confidence is calculated as the product of probabilities, $p_{tot}({\mathrm{T}}) =  \displaystyle\prod_{i=1}^n p(\tau_i)$.

	\item[Verbalized Self-Ask] \cite{linTeachingModelsExpress2022b, zhouNavigatingGreyArea2023b, tianJustAskCalibration2023} \emph{(Reflective, $p_v$)} We instruct the model to reflect jointly on the prompt, \emph{and} the model-generated code, and then output a numeric value of its confidence in the generated code.

	      Extra logic is implemented for when the model fails to output a probability (discussed further in~\cref{sec:verbalizeWacky}).

	\item[Question Answering Logit]~\cite{kadavathLanguageModelsMostly2022c} \emph{(Reflective, $p_B$ and $p_{NB}$)} In this case, rather than prompting for a numerical score (as above), we ask for a {\bf TRUE} or {\bf FALSE} answer. The probability associated with the {\bf TRUE} token is taken to be the confidence measure. Additionally, we extend this approach using \emph{normalization}: the model can assign probability mass to multiple possible expressions of {\bf TRUE} or {\bf FALSE} or even other variations (e.g, `` True'', `` true'', `` '', ). Thus when extending to the normalized form (Ask T/F N), we take the fraction of probability mass ``True'' assigned between only ``True'' and ``False''.
\end{description}

Our experiments consider the four above:
\emph{average token probability}, \emph{generated sequence probability}, \emph{verbalized self-evaluation},
and \emph{question answering logit}. In addition, as a baseline, we also used the \emph{length} of the generated
sequence. The length baseline is calculated based on the number of characters in the generated sequence, scaled such that $0$ is the shortest value in the dataset, and $1$ is the longest.
 
For reflective measures, we
expected that a code generation model should perform well (and provide well-calibrated confidence scores for correctness): first, \emph{for synthesis}, given just a good natural language description (without tests), of the desired function; second, \emph{for completion \& bug-fixing}, given the surrounding context. In both settings, we measure correctness using available hidden test cases. We also note that including hidden or failing tests in an LLM prompt is not common experimental practice for the tasks we analyze.~\cite{chenEvaluatingLargeLanguage2021b,austinProgramSynthesisLarge2021b,jiangImpactCodeLanguage2023b}.

\subsection{Measures of Calibration}
\label{sec:calibration_measures}
Using model's \emph{confidence} in its generated output, and a way of determining  \emph{correctness}, one can compute measures of calibration.
Calibration measures conceptually arise from the \emph{Reliability Plot},
which plots correctness \vs confidence

Two reliability plots illustrating this method are shown in~\autoref{fig:relDiagram}.~\autoref{fig:codexTokenLevelReliability} is for token-level code completions from \textsc{Codex}; it shows the observed proportion of exact-match correct tokens  (y-axis) \vs the predicted probability \ie confidence \emph{as per} the language model, based on bucketing observations
into subsets $S_1, S_2, \ldots S_n$. Here, there are $n=10$ buckets, equally spaced by confidence measure. Each bucket has an associated bar whose height
indicates the proportion (value $\in [0,1]$) of correct samples in the bucket. The closer the bars in each $S_i$ are to the diagonal line, the better the calibration.

\begin{figure}[htb]
	\centering
	\begin{subfigure}[b]{0.48\columnwidth} 
		\centering
		\includegraphics[width=0.8\textwidth]{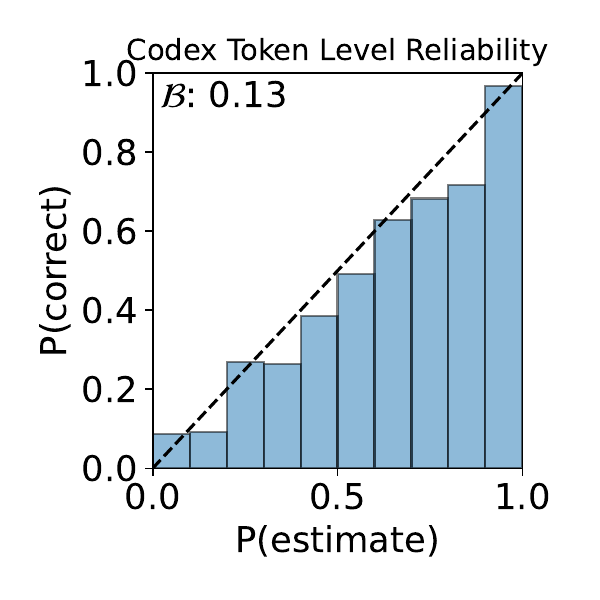}
		\caption{Well Calibrated}
		\label{fig:codexTokenLevelReliability}
	\end{subfigure}
	\hfill 
	\begin{subfigure}[b]{0.48\columnwidth} 
		\centering
		\includegraphics[width=0.8\textwidth]{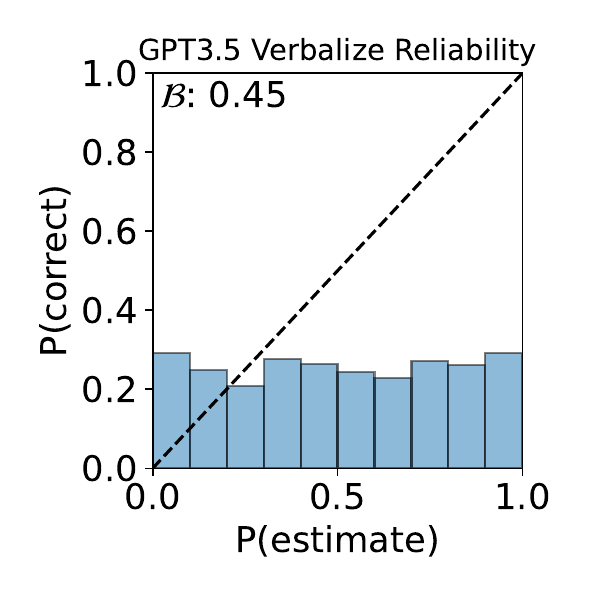}
		\caption{Poorly Calibrated}
		\label{fig:gpt35VerbalizedDyPyNoDocReliability}
	\end{subfigure}
	\caption{{\footnotesize\em Sample calibration plots demonstrating well- \emph{vs.} poorly- calibrated.}}
	\label{fig:relDiagram}
\end{figure}

We note here that \textsc{Codex} is a large model, well-trained on the task of token-level completion; thus, it is both well-calibrated and generally correct on the simple token-level completion task. However, for notions of correctness farther from the training objective, such as line-level code completion and test-passing, \textsc{Codex}'s \emph{intrinsic} probability may not be as well-calibrated. An example (\autoref{fig:gpt35VerbalizedDyPyNoDocReliability}) where the confidence is not well-calibrated is \gptturbo for line completion using verbalized confidence (\ie asking it to write its confidence; see \autoref{sec:confmeasure}).

We study two measures of calibration: Brier Score~\cite{brierVerificationForecastsExpressed1950a} and $\mathit{ECE}$~\cite{naeiniObtainingWellCalibrated2015a}. Both measure the deviation from
perfect calibration.
As before (following~\cite{guoCalibrationModernNeural2017b}), we assume a model $M$, input $X$, actual desired output $Y$, and model prediction $ M(X) = \hat{Y}$.
In our case, both $Y$ and $\hat{Y}$ are code, rather than a single label.
Calibration measures indicate the extent to which the deviations of $\hat{Y}$ from
the desired $Y$ are actually aligned with the model's confidence in its
output, $\hat{Y}$.

\useshortskip

From the calibration plot, with the evaluation
set $T$ bucketed into subsets
$S_i, i = 1 \ldots m \quad \text{s.t.}$ $\quad \bigcup_{i} S_i =  T, \quad \forall i \in 1 \ldots m$.
We estimate correctness in each bucket as the fraction of predictions in that bucket which are ``correct" (as discussed in~\sref{sec:correctness}); confidence is the average estimated probability from the model in that bucket:

\useshortskip
\begin{align}
	\text{corr}(S_i) & = \frac{1}{\mid S_i\mid} \sum_{x_i \in S_i} \mathds{1} ( \checkmark M(x_i)) \\
	\text{conf}(S_i) & = \frac{1}{\mid S_i\mid} \sum_{x_i \in S_i} \hat{p}_i
	\label{eqn:corrAndConf}
\end{align}
where the model generates the code $M(x_i)$ with confidence (probability) $\hat{p}_i$, and $\checkmark M(x_i)$ indicates
that the generated code is correct,
as per the operative experimental definition.
For a perfectly calibrated $M$, we would have $\text{corr}(S_i) = \text{conf} (S_i) \quad \forall i \in 1 \ldots m$. In practice,
we observe deviations from this ideal.
Expected Calibration Error ($\mathit{ECE}$)~\cite{naeiniObtainingWellCalibrated2015a} is a typical measure of calibration, calculated as the weighted average of these deviations.

\useshortskip
\begin{equation}
	\mathit{ECE} = \displaystyle\sum_{i=1}^m \frac{\mid S_i \mid}{\mid T \mid} \lvert \text{corr}(S_i) - \text{conf}(S_i) \rvert
	\label{eqn:ece}
\end{equation}

$\mathit{ECE}$ is intuitive, but can mislead; as seen below, a na\"ive predictor whose confidence is always the base rate would yield a deceptively low $\mathit{ECE}$ value.
An alternative measure, the Brier score, ${\mathcal B}$~\cite{brierVerificationForecastsExpressed1950a} avoids this issue; it
is calculated as follows:
\useshortskip
\begin{equation}
	\mathcal{B} = \frac{1}{\mid T \mid} \displaystyle\sum_{i=1}^{\mid T \mid} (\hat{p}_i - \mathds{1} (\checkmark M (x_i)))^2
	\label{eqn:brier}
\end{equation}
One can achieve an optimal Brier of $\mathcal{B} \approx 0$ when confidently estimating $\hat{p}_i \approx 1$ when the code is correct, and estimating $\hat{p}_i \approx 0$ when the code is incorrect for each sample.

For comparison, consider using the \emph{Unskilled Reference Brier Score,} $\mathcal{B}_{ref}$, attainable by a na\"ive,  ``unskilled'' model, which simply assigns the base-rate $p_{r}$ as its confidence for every prediction. Here, all prediction confidence values are in one bin, the value $p_r$; and the empirical correctness in this single bin \emph{is} the base rate $p_r$; so $\mathit{ECE} \approx 0$ which is misleading (thus exemplifying one of the weaknesses of $\mathit{ECE}$). The closed-form Brier Score for this unskilled predictor is:
\useshortskip
\begin{equation}
	\mathcal{B}_{ref} = p_r (1 - p_r)
	\label{eqn:brierRef}
\end{equation}

In a 50-50 coinflip scenario (assuming \textit{heads} is \textit{correct}), a na\"ive predictor that randomly guesses \textit{correct} with 50\% confidence receives $\mathcal{B}_{ref} = 0.5 * (1 - 0.5) \approx 0.25$. Higher base rates yield lower $\mathcal{B}_{ref}$; \eg for the \textsc{MBPP} dataset~\cite{austinProgramSynthesisLarge2021b}, \gptturbo generates test-passing solutions for about 72\% of the programming problems; here always guessing \textit{correct} with 72\% confidence results in $\mathcal{B}_{ref} = 0.72 * 0.28 \approx 0.20$. With well-calibrated confidence scores, a ``skilled'' model can achieve
Brier Scores lower than this unskilled $\mathcal{B}_{ref}$
value; if a model does worse, it is indicative of poor calibration.
Thus, one commonly reports a \emph{Skill Score} ($SS$), calculated thus:
\useshortskip
\begin{equation}
	SS = \frac{\mathcal{B}_{ref}-\mathcal{B}_{actual} }{\mathcal{B}_{ref}}
	\label{eqn:skillScore}
\end{equation}

Positive $SS$ (perfect score = 1.0) indicates improvement over baseline $\mathcal{B}_{ref}$;
negative indicates worse calibration than the baseline. Small positive values of $SS$ can sometimes indicate good skill. For example, the \emph{Deutsche Wetterdienst} (German weather forecasting service) considers 0.05 Skill Score to be a minimum threshold for a good forecast
quality\footnote{\href{https://www.dwd.de/EN/ourservices/seasonals_forecasts/forecast_reliability.html}{www.dwd.de/EN/ourservices/seasonals\_forecasts/forecast\_reliability.htm}}. As another data point, the American data journalism site \textit{538} reports a skill of around 0.13 in forecasting World Cup games\footnote{\href{https://projects.fivethirtyeight.com/checking-our-work/}{projects.fivethirtyeight.com/checking-our-work/}}, which is in the range of what we observe
in our experiments for best case code generation by LLMs; but these LLM performance numbers are just a starting point, and can be expected to improve in the future.
$\mathit{ECE}$ (\autoref{eqn:ece}) and Brier Score (\autoref{eqn:brier}) serve slightly different purposes: the Brier Score is calculated for each sample, and measures \emph{both} the ability to correctly discriminate output categories, \emph{and} calibration of the output probability. The $\mathit{ECE}$ measures just calibration; but it can be misleadingly low, as noted above for the unskilled predictor. Additionally binning must be done carefully, since it can affect $\mathit{ECE}$ scores~\cite{nixonMeasuringCalibrationDeep2019a}.

\subsection{Rescaling Approaches}
\label{sec:rescaledesc}
Machine learning models are not always calibrated.
\textcite{guoCalibrationModernNeural2017b} discuss ways of rescaling probability estimates to better match observations. A common approach is Platt scaling~\cite{plattProbabilisticOutputsSupport1999b}, where a logistic regression is fit to the logit values of the prediction \ie the $\ln$ of the measured confidence probability. This optimizes two parameters, a linear scaling multiplier and a bias \ie intercept, that shifts the value.

To reduce the likelihood that the scaling overfits
\& skews our results, we rescale over five folds; \ie we fit a logistic regression on a random \nicefrac{4}{5} of data and apply it to \nicefrac{1}{5} of data, before sliding over and doing each combination of \nicefrac{4}{5}.

Besides Platt scaling,
temperature rescaling has also been used~\cite{desaiCalibrationPretrainedTransformers2020b, kadavathLanguageModelsMostly2022c, guoCalibrationModernNeural2017b}: this approach applies a scalar multiplier on the logits representing each class \eg a multiclass image classifier. In our binary confidence case, this has similar expressivity to Platt scaling without an intercept. Other approaches include histogram binning \cite{zadroznyObtainingCalibratedProbability2001}, isotonic regression \cite{zadroznyTransformingClassifierScores2002}, \textit{inter alia}~\cite{guoCalibrationModernNeural2017b, kullTemperatureScalingObtaining2019}. These approaches are more parameterized; given the data limitations in our experimental setup \eg a few hundred examples in function synthesis, they pose higher risk of overfitting.

As we discuss in~\autoref{sec:result}, Platt scaling does improve calibration, with some caveats.

\subsection{Tasks \& Dataset}
\begin{table}[h]
	\centering
	\resizebox{1\columnwidth}{!}{
		\renewcommand{\arraystretch}{1.2}
		\begin{tabular}{|l|l|c|l|l|l|}
			\hline
			\multicolumn{1}{|c|}{Task}          & \multicolumn{1}{c|}{Dataset} & Dataset Size & \multicolumn{1}{c|}{\begin{tabular}[c]{@{}c@{}}Correctness \\ Measure\end{tabular}}      & \multicolumn{1}{c|}{Confidence Measure}                                                                                                                                              & \multicolumn{1}{c|}{\begin{tabular}[c]{@{}c@{}}Calibration \\ Metric\end{tabular}} \\ \hline
			\multirow{2}{*}{Function synthesis} & HumanEval                    & 164          & \multirow{2}{*}{\begin{tabular}[c]{@{}l@{}}Test-passing \\ Correctness\end{tabular}}     & \multirow{5}{*}{\begin{tabular}[c]{@{}l@{}}Average Token \\ Probability, Generated \\ Sequence Probability, \\ Verbalized Self-Evaluation, \\ Question Answering Logit\end{tabular}} & \multirow{5}{*}{\begin{tabular}[c]{@{}l@{}}Brier Score,\\ $ECE$\end{tabular}}      \\ \cline{2-3}
			                                    & MBBP Func                    & 880          &                                                                                          &                                                                                                                                                                                      &                                                                                    \\ \cline{1-4}
			Line-level Completion               & DyPyBench                    & 1,988        & \multirow{2}{*}{\begin{tabular}[c]{@{}l@{}}Test-passing \\ Correctness, EM\end{tabular}} &                                                                                                                                                                                      &                                                                                    \\ \cline{1-3}
			\multirow{2}{*}{Program Repair}     & Defects4J 1-line             & 120          &                                                                                          &                                                                                                                                                                                      &                                                                                    \\ \cline{2-4}
			                                    & ManySStubs4j                 & 3,000        & Exact-Match (EM)                                                                         &                                                                                                                                                                                      &                                                                                    \\ \hline
		\end{tabular}
	}
	\caption{List of tasks with associated datasets and measures.}
	\label{tbl:data}
\end{table}

\subsubsection{Function Synthesis}
This task aims to generate Python functions from ``Docstrings''.\footnote{Docstrings are code comments that explain the code's purpose and usage. Further discussed in~\sref{sec:dypymethod}.} Correctness is determined by functional testing.

We use \textsc{HumanEval}~\cite{chenEvaluatingLargeLanguage2021b} and \textsc{MBPP}~\cite{austinProgramSynthesisLarge2021b} datasets (see~\cref{fig:logprob} for sample prompts and model output).
One caveat: the samples in these datasets largely
constitute artificial problems, specifically assembled to test the code-synthesis capacity of LLMs;
measurements (both accuracy and calibration) over these datasets
may not generalize to real-world software development. Even so, these datasets provide a valuable datapoint for assessing model calibration.

We restructure all \textsc{MBPP} problems into a function synthesis task by placing the prompt inside the tested method as a Docstring,
making it comparable to \textsc{HumanEval}.
Additionally we exclude approximately 75 problems where the reference solution fails to pass the provided test cases\footnote{Due to either buggy reference code/tests, or possibly missing environment/networking/compute-time requirements.}.

\subsubsection{Line-level Code Completion}
\label{sec:dypymethod}
Code completion is
currently the most important and widely-deployed generative
task, with tools like GitHub Copilot~\cite{chenEvaluatingLargeLanguage2021b}.
Completion performance has been studied
at both the token and line levels~\cite{izadiCodeFillMultitokenCode2022a,kimCodePredictionFeeding2021b,luCodeXGLUEMachineLearning2021b,zieglerProductivityAssessmentNeural2022b}. However, \emph{calibration} for this vital, widely-deployed task
has so far not been evaluated in detail

The current decoder-only GPT models~\cite{brownLanguageModelsAre2020b,chenEvaluatingLargeLanguage2021b}
are \emph{already trained} to generate the next token at low average cross-entropy
given all the prior tokens, following the condition $p(token\ |\ prior\ tokens)$.
Unsurprisingly, we found such autoregressively-trained models are \emph{per se} well-calibrated at the token level. In this work, we will primarily focus on line-level completion. While several datasets exist for this problem~\cite{allamanisMiningSourceCode2013a,raychevProbabilisticModelCode2016b}, we use \textsc{DyPyBench}~\cite{islembouzeniaDyPyBenchBenchmarkExecutable2024}, a new dataset consisting of 50 popular open-source Python projects, including test suites for these projects. The test suites allow a test-correctness measure, in addition to the highly restrictive exact-match (\viz, the original line).

\textsc{DyPyBench} consists of complex real-world projects, each with hundreds of thousands of lines of Python code, and totaling over 2.2 million lines of Python code.
We ran all test suites for each project with coverage reporting enabled, extracted all functions from the projects, following~\cite{husainCodeSearchNetChallengeEvaluating2020a}, and selected 1,988 functions with at least 3 lines in the body, $100\%$ test coverage, and at least one line in the ``Docstring''.

\subsubsection{Program Repair}
Program repair is a well-studied problem in software engineering~\cite{gouesAutomatedProgramRepair2019a}. Several studies report that LLMs are effective at this task~\cite{fanAutomatedRepairPrograms2023b,jiangImpactCodeLanguage2023b,ahmedBetterPatchingUsing2023b}. However, LLM \emph{calibration} for program repair is not well-understood.
This paper focuses on small, pre-localized single-line bugs. We leverage the widely-used Defects4J dataset ~\cite{justDefects4JDatabaseExisting2014a}, which includes real-world examples of buggy programs, with fixes and test-sets. We extract 120 single-line bugs from \textsc{Defects4J} dataset. However, with only 120 samples, we may not obtain a comprehensive view of calibration. Therefore, we included another dataset, ManySStubs4J~\cite{karampatsisHowOftenSingleStatement2020a} (abbr. \textsc{SStubs}), which consists of single-line repairs. Following the setup of the \textsc{SStubs} dataset, the bug might be localized to a sub-expression of the line\footnote{Note, due to data processing errors, 3 Defects4J examples have slightly mis-localized bugs. We leave these as-is, reasoning that model confidence should be robustly well-calibrated even with slight localization noise, ideally giving a lower confidence of a fix if the location is noisy.}. We sample (uniformly at random) 3,000 examples from this dataset. \textsc{SStubs} does not provide test-sets; so the only evaluation metric available is the exact-matching of the generated text to the ground truth bug-free text.

\subsection{The Models}

We explore confidence calibration for three code generation models. These include OpenAI \gptturbo\footnote{The gpt-3.5-turbo-instruct model, \url{https://platform.openai.com/docs/models/gpt-3-5-turbo}}, OpenAI \textsc{Codex} \cite{chenEvaluatingLargeLanguage2021b}, and \textsc{CodeGen2-16B} \cite{nijkampCodeGen2LessonsTraining2023b}. We sample from the models with temperature of 0, consistent with the reality that busy developers typically look at just the first suggestion in the completion~\cite{barkeGroundedCopilotHow2023a}. For function synthesis task, temperature zero is most accurate and is fairly standard practice when doing pass@1 with only one solution to generate and present ~\cite{chenEvaluatingLargeLanguage2021b,liuYourCodeGenerated2023a}.

\section{Results}
\label{sec:result}

We begin with a brief overview of the findings on the correctness- \& confidence- measures of LLMs on the various tasks, and then provide detailed results on the calibration-related research questions.
\definecolor{improvement}{rgb}{1,0,0}
\definecolor{modelrow}{gray}{0.85}

\begin{table}[!h]
	\centering
	\resizebox{0.95\columnwidth}{!}{
		\begin{tblr}{
				colspec = {lcccc},
				vline{5} = {3-7}{gray!75,dotted},
				columns = {font=\small},
			}

			                           & \SetCell[c=3]{c} All Pass@1 &                  &                                    & \SetCell[c=3]{c} Exact-Match &                  &                  \\
			\cmidrule[lr,gray!75]{2-4} & CodeGen2                    & Codex            & \cmidrule[lr,gray!75]{5-7} GPT-3.5 & CodeGen2                     & Codex            & GPT-3.5          \\
			\midrule
			SStubs                     & -                           & -                & -                                  & 0.73\%                       & \textbf{27.77\%} & 20.27\%          \\
			DyPyBench                  & 28.84\%                     & 32.96\%          & \textbf{33.22\%}                   & 19.68\%                      & 23.60\%          & \textbf{23.96\%} \\
			Defects4J                  & 0.00\%                      & \textbf{23.33\%} & 19.17\%                            & 0.00\%                       & \textbf{19.17\%} & 15.00\%          \\
			HumanEval                  & 23.17\%                     & 47.24\%          & \textbf{64.60\%}                   & -                            & -                & -                \\
			MBPP                       & 29.08\%                     & 61.79\%          & \textbf{72.04\%}                   & -                            & -                & -                \\
			\bottomrule
		\end{tblr}
	}
	\caption{\scriptsize Performance comparison of models on tasks. Metrics are All Pass at Rank 1 (\textbf{All Pass@1}), meaning all project test cases passed with the line completion on first and only sample (at $t=0$), and Exact-Match, meaning the line completion was an exact string match with the original project line. Exact-Match is not commonly used for function synthesis tasks, since the generated output is longer and less likely to match. SStubs dataset does not have test cases. Boldface signifies high performing model for task and metric.}
	\label{table:modelPerformanceComparison}
	\vspace*{-2\baselineskip}
\end{table}
\subsection{RQ 1: How well are language models' confidence in their output aligned with the
	empirical correctness of their output, specifically for common generative tasks, \viz function synthesis, line-level code completion, and program repair?}

\subsubsection{Overall Correctness}
Correctness performance rate of the various models on the various tasks and datasets, are presented in~\autoref{table:modelPerformanceComparison}. Specifically, we report the fraction of samples passing all test cases for a given model and dataset, and the percentage of exact-matches. We found that \gptturbo worked well for both function synthesis (\textsc{HumanEval} and \textsc{MBPP}),
and line-level code completion,
whilst \textsc{Codex} generally performed well on program repair.
The \textsc{DyPyBench} benchmark reflects the most popular use of LLMs, \viz, for code completion.

\subsubsection{Correctness: Test-passing \vs Exact-match}
\label{sec:correct-descriptive}
As per~\sref{sec:correctness}, we evaluate correctness both on test-passing and exact-match.
Our experiment included two datasets (\textsc{Defects4J} and \textsc{DyPyBench}), where both methods of measuring correctness were available. Since \textsc{Defects4J} consists of only 120 samples,
we present the results for \textsc{DyPyBench}; in this case, as per~\autoref{table:modelPerformanceComparison}, \gptturbo performed best, with approximately 33\% of generated code passing all available tests, and approximately 24\% matching exactly.

In this setting (\textsc{DyPyBench}/\gptturbo), we cross-tabulate performance across the two correctness-measuring methods. We note that approximately {\bf half} the test-passing generations
did \emph{not} match the original code exactly; furthermore 6.89\% of the cases where the code matched exactly, did not pass
all the test cases. Upon careful study we found that these tests were ``flaky'', depending on network conditions, execution order, and other variable execution environment conditions. This aligns with~\cite{islembouzeniaDyPyBenchBenchmarkExecutable2024}, the author of this dataset, who observed an overall $7\%$ failure rate, but noted that 31 out of 50 projects had zero failed tests. \emph{This illustrates the relative merits/demerits of each correctness-evaluating approach, in practical SE settings.} Since the correctness performance is different with these two notions of correctness, the \emph{calibration} is also different, as we see below.

\subsubsection{Confidence Measures}
As might be expected, the two intrinsic measures $p_{avg}$ \& $p_{tot}$ are usually somewhat, and sometimes strongly, positively correlated with each other within the same model, dataset, and task.

\definecolor{improvement}{rgb}{1,0,0}
\definecolor{modelrow}{gray}{0.85}

\begin{table*}
    \centering
    \resizebox{0.8\textwidth}{!}{
        \begin{tblr}{
            colspec = {llccccccccccccccc},
            vline{3,6,9,12,15} = {4-24}{gray!75, dotted},
            columns = {font=\small},
                }
            &  & \SetCell[c=3]{c} \textbf{Line Completion} & & & \SetCell[c=6]{c} \textbf{Function Synthesis} & & & & & & \SetCell[c=6]{c} \textbf{Program Repair} & & & & &  \\

            \cmidrule[gray!75]{3-6} \cmidrule[lr,gray!75]{6-11} \cmidrule[l,gray!75]{12-17}
             & & \SetCell[c=3]{c} DyPyBench & & & \SetCell[c=3]{c} HumanEval & & & \SetCell[c=3]{c} MBPP & & & \SetCell[c=3]{c} Defects4J & & & \SetCell[c=3]{c} SStubs \\
            \cmidrule[lr]{3-5} \cmidrule[lr]{6-8} \cmidrule[lr]{9-11} \cmidrule[lr]{12-14} \cmidrule[l]{15-17}
            Model & Metric & ${\mathcal B}$ & $SS$ & $ECE$ & ${\mathcal B}$ & $SS$ & $ECE$ & ${\mathcal B}$ & $SS$ & $ECE$ & ${\mathcal B}$ & $SS$ & $ECE$ & ${\mathcal B}$ & $SS$ & $ECE$ \\
            \cmidrule[l]{1-17}
GPT-3.5
    & Total Prob
        & \textbf{0.23} & \textbf{-0.03} & 0.15
        & 0.62 & -1.70 & 0.63
        & 0.71 & -2.50 & 0.71
        & 0.25 & -0.63 & 0.28
        & 0.24 & -0.50 & 0.25
    \\
    & Avg Prob
        & 0.41 & -0.87 & 0.46
        & 0.27 & -0.18 & 0.23
        & \textbf{0.22} & \textbf{-0.09} & \textbf{0.14}
        & 0.68 & -3.39 & 0.73
        & 0.64 & -2.94 & 0.69
    \\
    & Ask T/F
        & 0.25 & -0.13 & 0.16
        & 0.34 & -0.47 & 0.37
        & 0.33 & -0.64 & 0.38
        & \textbf{0.15} & \textbf{+0.05} & \textbf{0.04}
        & \textbf{0.16} & \textbf{-0.01} & \textbf{0.04}
    \\
    & Ask T/F N
        & 0.25 & -0.15 & \textbf{0.15}
        & \textbf{0.23} & \textbf{+0.01} & 0.19
        & 0.22 & -0.11 & 0.16
        & 0.20 & -0.30 & 0.24
        & 0.22 & -0.34 & 0.23
    \\
    & Verbalize
        & 0.43 & -0.92 & 0.42
        & 0.28 & -0.24 & 0.22
        & 0.24 & -0.17 & 0.17
        & 0.58 & -2.72 & 0.60
        & 0.50 & -2.09 & 0.53
    \\
    & Length
        & 0.44 & -0.99 & 0.46
        & 0.23 & -0.03 & \textbf{0.15}
        & 0.22 & -0.10 & 0.16
        & 0.53 & -2.43 & 0.60
        & 0.53 & -2.26 & 0.60
    \\
    & Unskilled
        & 0.22 &  0.00 & 0.00
        & 0.23 &  0.00 & 0.00
        & 0.20 &  0.00 & 0.00
        & 0.15 &  0.00 & 0.00
        & 0.16 &  0.00 & 0.00
    \\
\hline[dashed]
Codex
    & Total Prob
        & \textbf{0.23} & \textbf{-0.02} & 0.16
        & 0.44 & -0.77 & 0.45
        & 0.60 & -1.52 & 0.60
        & 0.25 & -0.39 & 0.24
        & 0.20 &  0.00 & 0.09
    \\
    & Avg Prob
        & 0.46 & -1.07 & 0.50
        & 0.34 & -0.38 & 0.35
        & \textbf{0.24} & \textbf{-0.03} & \textbf{0.19}
        & 0.66 & -2.68 & 0.69
        & 0.58 & -1.90 & 0.62
    \\
    & Ask T/F
        & 0.24 & -0.09 & 0.12
        & 0.37 & -0.47 & 0.36
        & 0.49 & -1.06 & 0.50
        & \textbf{0.18} & \textbf{+0.01} & \textbf{0.07}
        & \textbf{0.19} & \textbf{+0.03} & \textbf{0.02}
    \\
    & Ask T/F N
        & 0.23 & -0.06 & \textbf{0.07}
        & \textbf{0.32} & \textbf{-0.29} & \textbf{0.30}
        & 0.42 & -0.79 & 0.43
        & 0.25 & -0.41 & 0.27
        & 0.23 & -0.14 & 0.18
    \\
    & Verbalize
        & 0.38 & -0.74 & 0.35
        & 0.42 & -0.67 & 0.40
        & 0.38 & -0.61 & 0.33
        & 0.47 & -1.65 & 0.50
        & 0.43 & -1.14 & 0.42
    \\
    & Length
        & 0.43 & -0.95 & 0.45
        & 0.44 & -0.77 & 0.44
        & 0.56 & -1.35 & 0.56
        & 0.50 & -1.79 & 0.55
        & 0.56 & -1.78 & 0.59
    \\
    & Unskilled
        & 0.22 &  0.00 & 0.00
        & 0.25 &  0.00 & 0.00
        & 0.24 &  0.00 & 0.00
        & 0.18 &  0.00 & 0.00
        & 0.20 &  0.00 & 0.00
    \\
\hline[dashed]
CodeGen2
    & Total Prob
        & \textbf{0.21} & \textbf{-0.02} & 0.15
        & \textbf{0.23} & \textbf{-0.30} & \textbf{0.23}
        & 0.29 & -0.41 & 0.29
        & - & - & -
        & - & - & -
    \\
    & Avg Prob
        & 0.44 & -1.16 & 0.50
        & 0.60 & -2.39 & 0.66
        & 0.58 & -1.80 & 0.61
        & - & - & -
        & - & - & -
    \\
    & Ask T/F
        & 0.23 & -0.10 & \textbf{0.14}
        & 0.25 & -0.38 & 0.24
        & \textbf{0.25} & \textbf{-0.19} & \textbf{0.21}
        & - & - & -
        & - & - & -
    \\
    & Ask T/F N
        & 0.33 & -0.59 & 0.35
        & 0.39 & -1.19 & 0.45
        & 0.39 & -0.88 & 0.43
        & - & - & -
        & - & - & -
    \\
    & Verbalize
        & 0.42 & -1.04 & 0.41
        & 0.43 & -1.39 & 0.42
        & 0.40 & -0.94 & 0.38
        & - & - & -
        & - & - & -
    \\
    & Length
        & 0.47 & -1.28 & 0.51
        & 0.38 & -1.14 & 0.42
        & 0.33 & -0.58 & 0.28
        & - & - & -
        & - & - & -
    \\
    & Unskilled
        & 0.21 &  0.00 & 0.00
        & 0.18 &  0.00 & 0.00
        & 0.21 &  0.00 & 0.00
        & - & - & -
        & - & - & -
    \\
\bottomrule
        \end{tblr}
    }
         \caption{\scriptsize Calibration measured as raw, non-scaled Brier Score (${\mathcal B}$, $\downarrow$ lower better), Skill Score ($SS$, $\uparrow$ higher better), and Expected Calibration Error ($ECE$, $\downarrow$ lower better), with respect to ``all passed'' notion of correctness, except SStubs which is ``exact-match''. CodeGen2 repair values are omitted as it does not perform the task with greater than 1\% accuracy. The ``Unskilled" row corresponds to a naive approach where the confidence is always returned as the base correctness rate, with Skill Score ($SS$) always zero by definition.}
         \label{tab:brierAllPassedRaw}
     \vspace*{-1\baselineskip}
\end{table*}

\subsubsection{Calibration without Rescaling} \autoref{tab:brierAllPassedRaw} presents the results for Line Completion, Function Synthesis, and Program Repair for each model and the raw confidence measure, without any rescaling method. We find raw confidence measures are poorly calibrated, with inconsistent exceptions.
In fact, the raw baseline rate (using the average fraction correct without considering the individual generation) is hard to beat; the best skill-score is around $0.05$.

For line completion, the $p_{tot}$ confidence measure is slightly worse than the baseline rate; calibration error is modest ($ECE \sim 0.15$).
The total probability improves on
the average probability, which is overconfident: the average token probability exceeds the $\sim30\%$ overall success rate.

For function synthesis with raw measures, $p_{tot}$ exhibits very poor calibration for \gptturbo and \textsc{Codex}, but not for \textsc{CodeGen2} on \textsc{HumanEval},
while the best intrinsic measure for MBPP is $p_{avg}$ for \gptturbo and \textsc{Codex}. The intrinsic measures are inconsistent; with average probability showing indicators of better calibration for \gptturbo and \textsc{Codex}, but not for \textsc{CodeGen2}.

For program repair, intrinsic measures are consistently below the base rate for both models and are as such, poorly calibrated. There are several caveats here. First, \textsc{Defects4J} is a small dataset, so findings may not generalize.
Second, \textsc{CodeGen2} performs poorly on \textsc{Defects4J}. Since \textsc{CodeGen2} is a smaller model without instruction tuning and relatively more limited reasoning capabilities
it gets ``distracted" by the buggy version shown in the prompt: it often just repeats the buggy lines. With very few correct outputs, the estimation of the confidence measure becomes unreliable. Therefore, we have removed the \textsc{CodeGen2} results from~\autoref{tab:brierAllPassedRaw}. A final caveat is that \textsc{SStubs} uses only exact-match as a correctness measure, which is quite different from a test passing measure.
\dispbox{In general, non-scaled confidence measures are only well calibrated for exact match on code completion; for test-passing correctness, they are poorly calibrated.}
\definecolor{improvement}{rgb}{1,0,0}
\definecolor{modelrow}{gray}{0.85}

\begin{table*}
    \centering
    \resizebox{0.8\textwidth}{!}{
        \begin{tblr}{
            colspec = {llccccccccccccccc},
            vline{3,6,9,12,15} = {4-24}{gray!75, dotted},
            columns = {font=\small},
                }
            &  & \SetCell[c=3]{c} \textbf{Line Completion} & & & \SetCell[c=6]{c} \textbf{Function Synthesis} & & & & & & \SetCell[c=6]{c} \textbf{Program Repair} & & & & &  \\

            \cmidrule[gray!75]{3-6} \cmidrule[lr,gray!75]{6-11} \cmidrule[l,gray!75]{12-17}
             & & \SetCell[c=3]{c} DyPyBench & & & \SetCell[c=3]{c} HumanEval & & & \SetCell[c=3]{c} MBPP & & & \SetCell[c=3]{c} Defects4J & & & \SetCell[c=3]{c} SStubs \\
            \cmidrule[lr]{3-5} \cmidrule[lr]{6-8} \cmidrule[lr]{9-11} \cmidrule[lr]{12-14} \cmidrule[l]{15-17}
            Model & Metric & ${\mathcal B}$ & $SS$ & $ECE$ & ${\mathcal B}$ & $SS$ & $ECE$ & ${\mathcal B}$ & $SS$ & $ECE$ & ${\mathcal B}$ & $SS$ & $ECE$ & ${\mathcal B}$ & $SS$ & $ECE$ \\
            \cmidrule[l]{1-17}
GPT-3.5
    & Total Prob
        & 0.21 & +0.07 & 0.03
        & \textbf{0.20} & \textbf{+0.15} & 0.09
        & 0.19 & +0.07 & 0.05
        & 0.16 & -0.05 &
        & \textbf{0.16} & \textbf{+0.03} &
    \\
    & Avg Prob
        & \textbf{0.20} & \textbf{+0.08} & 0.04
        & 0.23 & -0.02 &
        & 0.20 &  0.00 &
        & 0.16 & -0.03 &
        & 0.16 & +0.02 &
    \\
    & Ask T/F
        & 0.22 &  0.00 &
        & 0.20 & +0.12 & 0.11
        & 0.18 & +0.09 & 0.06
        & \textbf{0.15} & \textbf{+0.05} &
        & 0.16 &  0.00 &
    \\
    & Ask T/F N
        & 0.22 &  0.00 &
        & 0.20 & +0.14 & 0.07
        & \textbf{0.18} & \textbf{+0.11} & 0.04
        & 0.15 & +0.04 &
        & 0.16 & +0.01 &
    \\
    & Verbalize
        & 0.22 &  0.00 &
        & 0.24 & -0.05 &
        & 0.20 & +0.02 &
        & 0.17 & -0.09 &
        & 0.16 &  0.00 &
    \\
    & Length
        & 0.22 &  0.00 &
        & 0.24 & -0.03 &
        & 0.20 & +0.01 &
        & 0.16 & -0.06 &
        & 0.16 &  0.00 &
    \\
    & Unskilled
        & 0.22 &  0.00 &
        & 0.23 &  0.00 &
        & 0.20 &  0.00 &
        & 0.16 &  0.00 &
        & 0.16 &  0.00 &
    \\
\hline[dashed]
Codex
    & Total Prob
        & \textbf{0.20} & \textbf{+0.09} & 0.03
        & 0.22 & +0.11 & 0.08
        & 0.22 & +0.06 & 0.04
        & 0.18 & -0.01 &
        & \textbf{0.19} & \textbf{+0.05} & 0.02
    \\
    & Avg Prob
        & 0.20 & +0.09 & 0.04
        & \textbf{0.22} & \textbf{+0.14} & 0.07
        & \textbf{0.21} & \textbf{+0.12} & 0.06
        & 0.18 & -0.02 &
        & 0.19 & +0.05 & 0.02
    \\
    & Ask T/F
        & 0.22 &  0.00 &
        & 0.24 & +0.03 &
        & 0.24 &  0.00 &
        & 0.18 &  0.00 &
        & 0.19 & +0.04 &
    \\
    & Ask T/F N
        & 0.22 &  0.00 &
        & 0.24 & +0.03 &
        & 0.24 &  0.00 &
        & 0.18 & -0.01 &
        & 0.20 & +0.02 &
    \\
    & Verbalize
        & 0.22 &  0.00 &
        & 0.26 & -0.02 &
        & 0.24 & -0.01 &
        & \textbf{0.18} & \textbf{ 0.00} &
        & 0.20 &  0.00 &
    \\
    & Length
        & 0.22 & +0.01 &
        & 0.26 & -0.03 &
        & 0.24 & -0.01 &
        & 0.20 & -0.10 &
        & 0.20 &  0.00 &
    \\
    & Unskilled
        & 0.22 &  0.00 &
        & 0.25 &  0.00 &
        & 0.24 &  0.00 &
        & 0.18 &  0.00 &
        & 0.20 &  0.00 &
    \\
\hline[dashed]
CodeGen2
    & Total Prob
        & \textbf{0.19} & \textbf{+0.08} & 0.04
        & 0.18 &  0.00 &
        & 0.21 &  0.00 &
        & - & - &
        & - & - &
    \\
    & Avg Prob
        & 0.19 & +0.07 & 0.02
        & 0.17 & +0.03 &
        & 0.21 &  0.00 &
        & - & - &
        & - & - &
    \\
    & Ask T/F
        & 0.21 &  0.00 &
        & 0.18 & -0.01 &
        & \textbf{0.20} & \textbf{+0.01} &
        & - & - &
        & - & - &
    \\
    & Ask T/F N
        & 0.21 &  0.00 &
        & \textbf{0.17} & \textbf{+0.04} &
        & 0.21 &  0.00 &
        & - & - &
        & - & - &
    \\
    & Verbalize
        & 0.21 &  0.00 &
        & 0.18 & -0.01 &
        & 0.21 &  0.00 &
        & - & - &
        & - & - &
    \\
    & Length
        & 0.21 &  0.00 &
        & 0.18 & -0.02 &
        & 0.21 &  0.00 &
        & - & - &
        & - & - &
    \\
    & Unskilled
        & 0.21 &  0.00 &
        & 0.18 &  0.00 &
        & 0.21 &  0.00 &
        & - & - &
        & - & - &
    \\
\bottomrule
        \end{tblr}
    }
         \caption{\scriptsize Calibration measured as Platt-scaled Brier Score (${\mathcal B}$, $\downarrow$ lower better), Skill Score ($SS$, $\uparrow$ higher better), and Expected Calibration Error ($ECE$, $\downarrow$ lower better), with respect to ``all passed'' notion of correctness, except SStubs which is ``exact-match''. In cases where the $SS$ is less than 0.05, the $ECE$ is omitted. This is because an estimate without any signal will become Platt-scaled to approximately the base rate. This will \emph{appear} as one well calibrated bin, resulting in an $ECE$ near zero, but does not provide information. CodeGen2 repair values are omitted as it does not perform the task with greater than 1\% accuracy.}
         \label{tab:brierAllPassedScaled}
     \vspace*{-1\baselineskip}
\end{table*}
\subsection{RQ 2: Can alignment between LLM confidence in generated code, and its correctness, be improved by confidence rescaling?}
\label{sec:rescalingresults}
\autoref{tab:brierAllPassedScaled} shows the results after applying Platt scaling to all measures (See \sref{sec:rescaledesc}). \autoref{fig:scalecompare} shows a reliability plot before rescaling, and its equivalent after rescaling in~\autoref{fig:dypyrely}.
Rescaling can improve calibration. Considering all values, $ECE$ improves from an average of \AverageEceNoScale to \AverageEceScaled. For just those measures with a post-scaling $SS$ of at least 0.05, $ECE$ improves from \AverageEceNoScaleMinSkill to \AverageEceScaledMinSkill.

\begin{figure}[htb]
	\centering
	\begin{subfigure}[b]{0.48\columnwidth}
		\centering
		\includegraphics[width=0.9\textwidth]{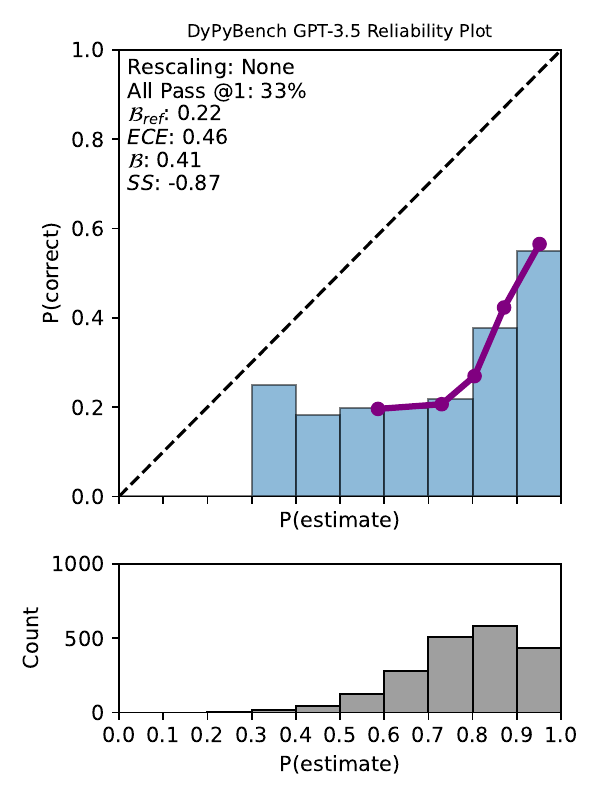}
		\caption{DyPyBench, nonscaled reliability plot}
		\label{fig:scalecompare}
	\end{subfigure}
	\hfill
	\begin{subfigure}[b]{0.48\columnwidth}
		\centering
		\includegraphics[width=0.9\textwidth]{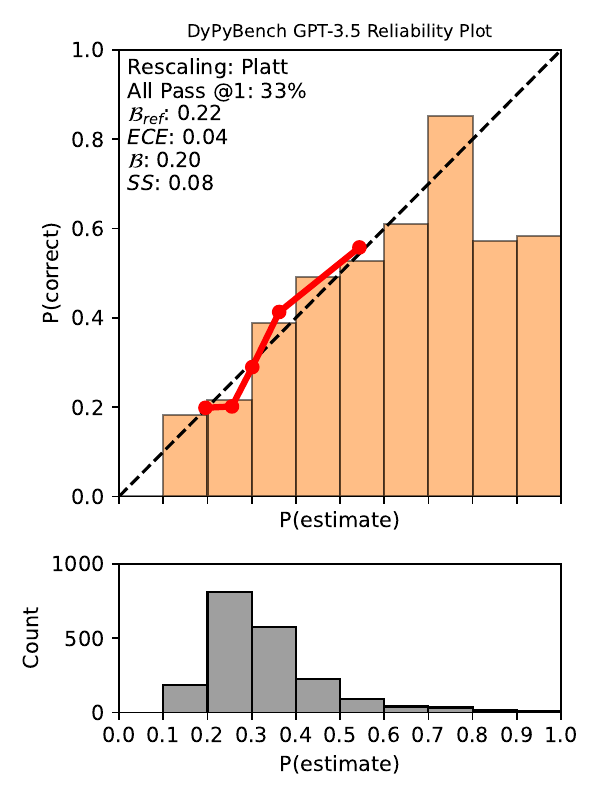}
		\caption{DyPyBench, Platt scaled reliability plot}
		\label{fig:dypyrely}
	\end{subfigure}
	\caption{\scriptsize Reliability plots for \textsc{DyPyBench} line-level code completion tasks, with respect to All Pass @1 correctness measure and Average Token Probability confidence measure. \gptturbo was used for both experiments. Bottom histogram represents number of samples in each bin. $\mathcal{B}_{ref}$ refers to the unskilled predictor Brier, $ECE$ to Expected Calibration Error, $\mathcal{B}$ to Brier Score, and $SS$ to Skill Score. Red \& purple lines represent scaled \& non-scaled quantile bins rather than evenly spaced bins with 1/5 of the data at each point. The left nonscaled plot shows over-confidence, as the confidence estimate is high, but the actual correctness is low. The scaled plot (right) improves calibration.}
    \label{fig:dypybenchReliabilityAndROC}
\end{figure}

\paragraph{Understanding ``Bucket Collapse''}\label{para:bucketCollapse}  Platt scaling can lead to deceptively low $ECE$.
If a confidence measure is poorly aligned with correctness, Platt-scaling
can rescale (squash) all the confidence values to the baseline rate;
this places all samples
in a single confidence value bucket where probability exactly matches the baseline rate of correctness, resulting in an $ECE$ near 0. This indicates the problem of only considering $ECE$.
$SS$ and Brier, on the other hand, would reveal
the poor utility of the confidence measure.
Thus when applying rescaling, it is important to consider Skill Score, rather than only Brier and $ECE$.

\paragraph{Results} After  rescaling, only the intrinsic measures show skill improvement over the baseline rate for line completion. $p_{avg}$ and $p_{tot}$ are similarly calibrated. The calibration and skill appears roughly consistent between all three models in this case.
Rescaling improves calibration results for function synthesis. The $p_{tot}$ measure reaches a $SS$ of 0.15 for \textsc{HumanEval}.
Rescaling useful improvement for reflective prompts as well bringing $SS$ and $ECE$ to similar values (discussed further in \autoref{sec:reflectionrq}).
For program repair, rescaling doesn't improve skill score, for any measure.

\paragraph{Is Rescaling a Panacea for Calibration?}
Rescaling typically improves calibration; it has been used in settings other than generative models of code, with other notions of correctness ~\cite{guoCalibrationModernNeural2017b,mindererRevisitingCalibrationModern2021a,desaiCalibrationPretrainedTransformers2020b,parkCalibrationPretrainedLanguage2022a,chenCloseLookCalibration2023b,bommasaniHolisticEvaluationLanguage2023,liOperationalCalibrationDebugging2020a}. However, there are disadvantages. First, ``bucket collapse'' (see ~\sref{para:bucketCollapse}) can mislead with deceptively low $ECE$. Second, some
correctness data is needed to fit rescaling parameters. When sweeping through various sized bootstrapped subsets of the data, we find that it can take over 64 data points for the rescaling to result in positive skill and lower $ECE$, with improvements continuing into 100s of data points (see~\cref{fig:bootstrapRescale} for bootstrapping analysis).
When using the full data, the rescaling between tasks can vary dramatically.\footnote{As seen in~\cref{fig:rescalingFacet}, which shows the curves learned between measure-task pairs.}
Ideally, we want confidence measures which are reliable and allow trustworthy auditing even when applying language models to new software engineering tasks. To study how close we are to this, we fit rescaling parameters to one task, and then apply it to the other tasks (see~\cref{fig:reuseMatrix}).
We find it is viable to use rescaling between tasks of the same domain with similar base rates, such as within the program synthesis tasks. For example, for \gptturbo when fitting $p_{NB}$ (results in next section) rescaling to each of two function synthesis tasks, and then applying it to the other, we observe an average drop of $SS$ from 0.14 $\rightarrow$ 0.12, and average drop of $ECE$ from 0.07 $\rightarrow$ 0.05. However, applying the $p_{total}$ rescaling fit on \textsc{DyPyBench} to the function synthesis tasks, results in an average $SS$ change of 0.12 $\rightarrow$ -1.28 and $ECE$ change of 0.08 $\rightarrow$ 0.54, indicating a lack of robustness.
These reasons suggest one must be careful when analyzing and reporting calibration results based on rescaling, and highlights the need for further work on confidence measures that might be more directly calibrated.

Without rescaling, total probability $p_{\text{total}}$ shows hints of calibration. With rescaling, there is a possible 10-20\% improvement over baseline rates and good calibration, but it is inconsistent as skill is poor for \textsc{CodeGen2} on Function Synthesis.

\dispbox{Rescaling is an effective technique for improving calibration, but metric improvements (${\mathcal B}$ and $ECE$) may be misleading by matching the base rate in a ``bucket collapse'' scenario or can lack generalization.}

\subsection{RQ 3: Is confidence obtained by reflection better aligned with correctness?}\label{sec:reflectionrq}
The two logit-based reflective measures $p_B$ \& $p_{NB}$ are usually strongly positively correlated with one another, since they are calculated from similar numbers. The reflective verbalized self-ask confidence measure $p_v$, and the two logit-based reflective confidence measures have no consistent relationships.

For function synthesis with raw measures, $p_{NB}$ shows best calibration for \gptturbo, slightly better than Unskilled, for \textsc{HumanEval}. For program repair, we observe the strongest best-case performance with regards to the metrics. Both \gptturbo and \textsc{Codex} show positive $SS$ and low $ECE$ for $p_B$, but they are inconsistent after normalization ($p_{NB}$). These metrics suggest with reflection, these models' confidence is calibrated regarding repair correctness; however,
further analysis (see~\cref{postageplot})
does not indicate good calibration on this task, from any confidence measure. We find that in general, the intrinsic \vs reflective measure values have no consistent relationship, even for a given model, dataset and task.

This lack of relationship may not necessarily be negative: \eg
perhaps the model's reflective, prompted confidence may be better calibrated, as suggested by prior work~\cite{baiConstitutionalAIHarmlessness2022b, austinProgramSynthesisLarge2021b}.
Without rescaling or few-shot prompting, reflective results are inconsistent. In some cases, such as \gptturbo \textsc{HumanEval} and \textsc{Defects4J}, there are signs of calibration with slightly positive $SS$ values and $ECE$ values less than 0.2. Normalizing the T/F values induces some difference; but there are inconsistencies \emph{vis-\`a-vis} tasks and models.
For nonscaled \gptturbo results, $p_{NB}$ improves calibration in line completion and function synthesis by an average of \GptNNTfNormalizationDeltaRawSs $SS$ and \GptNNTfNormalizationDeltaRawEce $ECE$, but not for program repair or for \textsc{CodeGen2}. Rescaling generally removes any sign of a normalization trend. For the alternative reflective approach of verbalization, the probability is not well calibrated for these models on the studied SE tasks.

In some cases, the reflective approaches are best calibrated without rescaling (see \autoref{tab:brierAllPassedRaw}), and show signs of being more robust when reusing learned rescaling parameters on unseen tasks (see~\cref{fig:reuseMatrix}).

\dispbox{Reflective approaches may be best calibrated ``out-of-the-box'' and in some settings when rescaled. However, in the tested settings, they do not improve significantly over intrinsic measures.}

\subsection{RQ 4: Can we use few-shot techniques to achieve better calibrated confidence for code completion, using an instruction-tuned model with in-context learning?}
\label{sec:rq4}
We investigated the impact of few-shotting \viz providing a model completion and correctness as part of the $p_{NB}$ prompt, on calibration~\cite{kadavathLanguageModelsMostly2022c}. To effectively perform few-shotting, we need a model that is instruction tuned and sufficiently large, which is best matched by \gptturbo.
We explore few-shotting for the widely-used line completion task.

We perform the experiment with 5-shots consisting of prior completions from the same experiments presented in \autoref{tab:brierAllPassedScaled}, the reflective question, and the ground truth True/False. We try two variants of this experiment, one where the examples are randomly selected, and one where they are chosen based on the similarity to the unanswered prompt. In both cases, we exclude the ground truth result for the unanswered prompt. We focus on Line Completion for this experiment as it represents widespread use and has a large number of examples available.

\begin{table}[!h]
	\centering
	\resizebox{0.8\columnwidth}{!}{
		\begin{tblr}{
				colspec = {lccc},
				columns = {font=\small},
			}

			Confidence Measure                & ${\mathcal B} \downarrow$ & $SS \uparrow$ & $ECE \downarrow$ \\
			\cmidrule[l]{1-11} 0-Shot Reflect & 0.25                      & -0.15         & 0.15
			\\
			0-Shot Reflect (Rescaled)
			                                  & 0.22                      & 0.00          &                  \\
			\cmidrule[l]{1-11}
			FS Random                         & 0.29                      & -0.29         & 0.21             \\
			FS Random (Rescaled)              & 0.22                      & 0.0           &                  \\
			FS BM25                           & 0.20                      & 0.08          & 0.10             \\
			FS BM25 (Rescaled)                & 0.19                      & 0.15          & 0.02             \\
			\bottomrule
		\end{tblr}
	}

	\caption{\scriptsize Few-shot reflective prompting using \gptturbo for line completion. `FS Random' refers to selecting random few-shot examples. `FS BM25' retrieves more relevant known completions. ECE values when rescaled values SS are close to zero are omitted (to avoid confusion with ``bucket collapse'', \autoref{para:bucketCollapse})}
	\label{table:fewshot}

\end{table}

For line completion, the non-scaled results using random examples did not result in improved calibration over the baseline $p_{NB}$, however using BM25~\cite{bm25fewshot} to select similar examples yielded a positive $SS$ of 0.08, which could be improved further by rescaling, up to 0.15. This result notably exceeds any other measure for \textsc{DyPyBench}, and significantly improves over the baseline $p_{NB}$ $SS$ of 0.

While random few-shotting requires limited extra data, BM25 is more similar to rescaling, in that it is dependent on a larger set of ground truths. This could be actualized by logging user completions, and building up ground truths (on if the completion was correct) based off the test case runs or acceptance of completions.

\dispbox{Providing a few examples selected via BM25 when asking \gptturbo to reflect on its own completion output significantly improves reflective calibration.}

Alternative and improved ways of prompting (\eg different verbalization formats \cite{linTeachingModelsExpress2022b, tianJustAskCalibration2023}, fine-tuning \cite{kadavathLanguageModelsMostly2022c, linTeachingModelsExpress2022b}, chain-of-thought \cite{weiChainofThoughtPromptingElicits2022, tianJustAskCalibration2023}, \etc ) may alter these findings and are areas for future work.

\section{Discussion}
\label{sec:discussion}
Language models are now widely-integrated into Software Engineering
practice, via tools like Copilot~\cite{zieglerProductivityAssessmentNeural2022b} and Didact\footnote{\href{https://blog.research.google/2023/05/large-sequence-models-for-software.html}{blog.research.google/2023/05/large-sequence-models-for-software.html}}.
We raise here the importance of calibration when integrating \emph{generative}
LLMs in coding practice. We evaluate the calibration of generative LLM use (especially code completion) with large samples of \emph{realistic} data (\textsc{DyPyBench}, \textsc{SStubs}), using widely adopted models, as well as some more academic datasets (\textsc{HumanEval}, \textsc{MBPP}).

\paragraph{Using a well-calibrated model--beyond simple defect prediction}
\label{par:test}
To clarify how a well-calibrated model enables
more well-grounded decision-making concerning generated outputs,
as compared to as compared to a traditional process choosing a binary decision point---we
consider \gptturbo working on
code completion, where
correctness is determined by test-passing, and confidence is assigned by few-shotting, average token probability. The base correctness (test-passing) rate of completions is about 33\%. 
With few-shotting, we get a very high skill score
of 0.15 (\autoref{sec:rq4}, \autoref{table:fewshot}). 

\begin{figure}[htb]
	\centering
	\includegraphics[width=0.5\columnwidth]{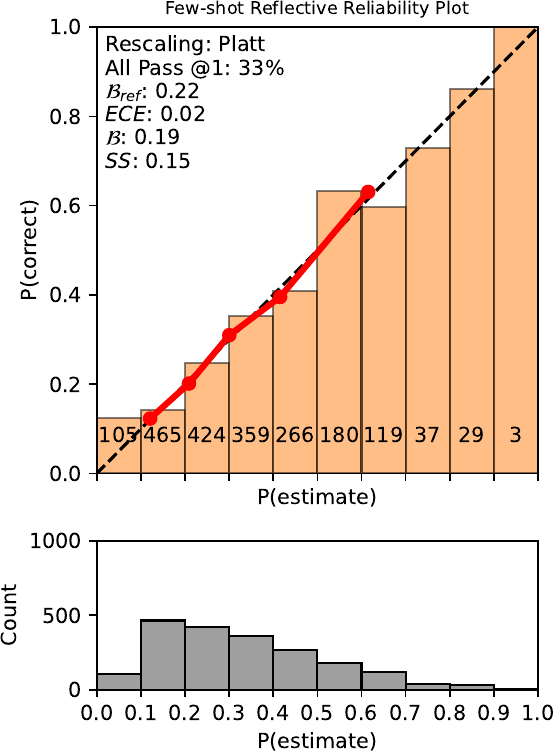}
	\caption{{\footnotesize\em Few-shot reflective reliability plot, based on ``FS BM25" row of \autoref{table:fewshot}}}
	\label{fig:fewshotrely}
\end{figure}

If we didn't have a well-calibrated model, 
we might very cautiously accept only those 
completions that are generated at a very high-confidence threshold; here, the FP rate could be low
(of course TP rate would be low as well). While this may lower the risk of bad code, it also regrettably reduces the available help from the LLM. However, a well-calibrated confidence measure allows a more rational, graduated set of decisions. Such a well-calibrated measure is visualized in \autoref{fig:fewshotrely}. In this setting, for much of the confidence scale, a user could look at the confidence level, and get a very good idea of how likely the code is to be correct, and make a well-reasoned, situation-specific set of decisions 
to manage risk, and allocate reviewing resources,
based on the model's confidence.
This provides an illustration of the greater benefit provided a well-calibrated measure (high Skill level, low ECE) over one that is just providing good precision-recall trade-off (or, ROC curve);
the latter does not
allow such graduated deployment of quality-control effort.
However, developers would need to learn to use calibrated probabilities in decision-making.

\paragraph{Beyond simple ``correctness''} In addition to the above uses, which considered a single notion of ``correctness'', one could consider a multi-class correctness prediction task, where the model could indicate the confidence in correctness (the absence of defects) from multiple perspectives: severity of possible defect, the kind of defect (relating to security, integrity, privacy, fairness, \etc) and defect complexity (indicating the cost or schedule impact of repairs). Drawing an analogy to classical forecasting, this is analogous to not just the probability it will rain, but probability it will be a drizzle or be a drenching thunderstorm.

\paragraph{Why calibration now?}
We've always had bugs; poor-quality code isn't
new. Our push for calibration, however,
arises from the increasing amount of code generated by LLM. GitHub
claims that up to 61\% of code\footnote{\href{https://github.blog/2023-02-14-github-copilot-now-has-a-better-ai-model-and-new-capabilities/}{github.blog/2023-02-14-github-copilot-now-has-a-better-ai-model-and-new-capabilities}} in some systems is generated by LLMs. It is also known that LLMs make a lot of mistakes. A recent paper has reported~\cite{jesseLargeLanguageModels2023b}
that LLMs, even when trained on \emph{properly fixed} code, tends to recapitulate
the \emph{the old unfixed, buggy code} when prompted in context. However, LLMs do have very
high capacity, and a demonstrated ability to usefully reflect~\cite{shinnReflexionLanguageAgents2023b,baiConstitutionalAIHarmlessness2022b} on their generated text. Thus,
we have both a high risk (of buggy code), and a chance to improve productivity.
We believe that improved calibration could lead to better management of the risk-benefit
of LLM-generated code. The studied correctness calibration is a stepping stone for more complex notions of confidence (like severity and localized confidence). Additionally, by studying code LLMs, we might make progress on the general safe deployment of capable generative models \cite{grosAISafetySubproblems2023b}.
\paragraph{Summarizing per-token probabilities}

To produce a summary confidence for generated token sequences in ~\cref{tab:brierAllPassedRaw,tab:brierAllPassedScaled}, in Tables \ref{tab:brierAllPassedRaw} \& \ref{tab:brierAllPassedScaled}, we used (arithmetic mean) average \& product  to summarize
the per-token probabilities. In this setting, 
it might be more reasonable to use geometric mean (as in \cite{Liu2023LitCabLL}) to 
get a product value normalized for length;
indeed, when we tried that, we found that Brier and skill scores improved marginally, but consistently so; future
research could indicate whether these findings
generalize. 

\section{Threats to Validity}
\paragraph{Sample Size \& Generalizability}
While three of our datasets contain more than 800+ samples each, \textsc{HumanEval} and \textsc{Defects4J} datasets consist of only 164 and 120 samples, respectively.
Results on these datasets may not generalize. However, we note that our study
has a large and natural dataset for the \emph{line-level code completion} task, which has current
practical importance. Given the noise and variance we observe, we recommend future work push towards larger and more natural datasets, in particular for Function Synthesis \& Repair.

While some of our treatments (\eg few-shotting) suggest substantial improvements in skill score (\autoref{table:fewshot}), in other cases such as the different approaches to summarize per-token confidence differences, the differences are less clear. In future work, these differences could be judged more robustly using bootstrapped p-values and effect sizes.

\paragraph{Artificial \vs real world data}
For function synthesis, we used the popular \textsc{HumanEval} and \textsc{MBPP} function synthesis datasets. These datasets contain small-ish Python programs that may not represent real-world software development functions. However, our other datasets, such as \textsc{DyPyBench} and \textsc{SStubs} are more representative of real-world, open-source GitHub projects.

\paragraph{Model Selection} Results might not generalize to all models, especially those with greatly differing training/finetuning or different architectures.

\paragraph{Experimental Design}
Our exploration is not exhaustive; other SE tasks and datasets also could benefit from calibration studies. Additionally, the specific prompts we used for this paper surely played a role in our findings. Other prompts or problem phrasings (such as different forms of context for line-level code completion) may yield different results. Regarding
test flakiness: the test ``flake'' rate of DyPyBench is not zero, but is quite low and
not unrealistic~\cite{luoEmpiricalAnalysisFlaky2014a}.

\vspace{.1in}
Despite these caveats, our study, which includes three tasks and five datasets, provides
a good starting point for further studies.

\section{Related Work}

LLMs for code are extensively studied~\cite{zhangSurveyLearningbasedAutomated2023a,zhengSurveyLargeLanguage2024}. While calibration has a long history in modeling~\cite{brierVerificationForecastsExpressed1950a, steyerbergAssessingPerformancePrediction2010a}, it is not a frequently studied topic in the SE community. Early work moving into modern machine learning studied the calibration of smaller neural
models performing classification tasks on text and images; while these early models were poorly calibrated \emph{per se}, their performance could be improved by simple scaling~\cite{guoCalibrationModernNeural2017b} of their output probabilities. As models became larger, calibration was found to improve~\cite{srivastavaImitationGameQuantifying2023b}. Pre-training was also found to improve calibration~\cite{hendrycksUsingPreTrainingCan2019b,desaiCalibrationPretrainedTransformers2020b}; however, these findings have been disputed~\cite{chenCloseLookCalibration2023b}.

More recent works evaluated LLM calibration on a wide variety of settings~\cite{kadavathLanguageModelsMostly2022c,jiangHowCanWe2021a,desaiCalibrationPretrainedTransformers2020b,keyTrustworthyNeuralProgram2023}. ~\citet{desaiCalibrationPretrainedTransformers2020b} studied non-code (natural language) tasks such as inference or paraphrasing, with only intrinsic measures using older-generation models (BERT and RoBERTA).~\citet{jiangHowCanWe2021a} studied calibration for natural language question-answering using just intrinsic measures. In contrast, we study calibration for three coding-related tasks, using both artificial and natural code datasets, and both intrinsic and reflective confidence measures, to evaluate calibration in the SE domain.

Other prior work has investigated tokens that might be edited. \citet{vasconcelos2022generation} discusses code model uncertainty for function-synthesis-style problems, and ran human evaluation of the usefullness of colored highlighting of uncertain tokens. They found highlighting a human-derived ground-truth of which tokens might be edited was helpful, and more useful than raw token probabilities from the model. \citet{rusure} developed method of highlighting likely edit tokens via a utility optimization algorithm comparing different file completions. We find exploring more on calibrated uncertainty for local areas be a interesting area for additional work.

\citet{liOperationalCalibrationDebugging2020a} investigate the calibration of Computer vision (CV) models from an operational perspective \ie the shift between training input and production inputs, presenting it as a software quality problem that can be addressed using Bayesian approaches. \citet{mindererRevisitingCalibrationModern2021a} evaluate the calibration of at the time, state of the art CV models and find improved calibration with more recent models, notably those not using convolutions. \citet{parkCalibrationPretrainedLanguage2022a} study the effect of the mixup technique~\cite{zhangMixupEmpiricalRisk2018a} on calibration in a natural language understanding (NLU) setting using older generation models (BERT and RoBERTa). \citet{chenCloseLookCalibration2023b} investigate the calibration of pretrained language models on various NLP tasks, also using older generation models (RoBERTa and T5).
\citet{bommasaniHolisticEvaluationLanguage2023} introduce the HELM benchmark, which includes calibration as one of its seven metrics to evaluate language models in a natural language context. \citet{lookleap} explored LM uncertainty with a range of techniques and tasks, including both NLP and function synthesis tasks. They evaluated using correlation measures, rather than focusing on calibration. They explore interesting sample-based and perturbation techniques which could be explored more for calibration on diverse SE tasks. Other work \cite{lever} has explored training an ML model that sees code and execution results to estimate correctness probabilities for solution reranking. For natural language question answering tasks, work has explored improving calibration by training a model to adjust token logits \cite{Liu2023LitCabLL}, and training a model from LLM hidden states specifically around the $ECE$ metric \cite{Liu2024EnhancingLM}.

When suitably prompted, \citet{kadavathLanguageModelsMostly2022c} found that LLMs can output well-calibrated scores on whether their own answers are correct or not, \viz, larger models ``know what they know''. While this work did investigate some function synthesis tasks (\textsc{HumanEval} \& an unpublished Python function dataset), they did so using only their private models, and ultimately focused on natural language tasks.~\citet{keyTrustworthyNeuralProgram2023} developed an approach that given a natural language problem description, produces a confidence score for a sampled candidate solution based on generated specifications, allowing them to judge whether the LLM can solve the problem at all. Their metrics include calibration. Recent work has also explored calibration of software topics such as root cause analysis~\cite{zhangPACELMPromptingAugmentation2023b}.

\section{Conclusion}
In this paper, we begin with the observation that while LLMs
are often helpful (for example producing code-completions
for developers) they often produce
buggy code. We
argue that a \emph{well-calibrated} confidence score,
could provide a reliable indication of whether the generated code
was correct,
and help more rational, graduated quality-control of
of LLM-generated code
We studied the calibration of intrinsic and reflective confidence measures in several practical settings (completion and repair) and a widely-used competitive setting (synthesis),  across several LLMs.
We find that LLMs are generally poorly calibrated
out of the box,
across a variety of confidence measures (both intrinsic and reflective)
We then found that Platt scaling generally results in somewhat better calibrated confidence measures.

Finally, we focused in on a) coding task where LLMs are most widely-deployed,
\viz code completion, and b) a very widely used instruction-tuned model, \viz \gptturbo, and investigated whether a reflective,  in-context learning approach (few-shotting) could provide better calibrated confidence measures. In this setting, we found that calibration improves substantially, reaching a skill score of $0.15$, particularly with retrieval augmented few-shotting.

To our knowledge, our paper is the first to consider the problem of calibration
in a real-world code generation setting. We do find that most models, both out-of-the-box and with simple reflection, don't provide reliable confidence measures. However, our results with retrieval-augmented few-shotting are very encouraging, and point towards a future where Language Models could provide developers with guidance on how to quality-control the code they generate.

\section{Acknowledgments}
We acknowledge partial support for this work by
the Intelligence Advanced Research Projects Agency (IARPA) under contract W911NF20C0038,
the National Science Foundation under CISE SHF MEDIUM 2107592,
the European Research Council (ERC grant agreement 851895), and
the German Research Foundation (ConcSys, DeMoCo, and QPTest projects). Devanbu was supported by a Humboldt Research Award\footnote{\url{https://www.humboldt-foundation.de/en/connect/explore-the-humboldt-network/singleview/1226147/prof-dr-premkumar-t-devanbu}}.  
Our conclusions do not necessarily reflect the position or the policy of
our sponsors and no official endorsement should be inferred.

\printbibliography

@inproceedings{ahmedBetterPatchingUsing2023b,
  title = {Better {{Patching Using LLM Prompting}}, via {{Self-Consistency}}},
  booktitle = {2023 38th {{IEEE}}/{{ACM International Conference}} on {{Automated Software Engineering}} ({{ASE}})},
  author = {Ahmed, Toufique and Devanbu, Premkumar},
  date = {2023-09},
  pages = {1742--1746},
  issn = {2643-1572},
  eventtitle = {2023 38th {{IEEE}}/{{ACM International Conference}} on {{Automated Software Engineering}} ({{ASE}})}
}

@inproceedings{allamanisMiningSourceCode2013a,
  title = {Mining Source Code Repositories at Massive Scale Using Language Modeling},
  booktitle = {2013 10th {{Working Conference}} on {{Mining Software Repositories}} ({{MSR}})},
  author = {Allamanis, Miltiadis and Sutton, Charles},
  date = {2013-05},
  pages = {207--216},
  issn = {2160-1860},
  eventtitle = {2013 10th {{Working Conference}} on {{Mining Software Repositories}} ({{MSR}})}
}

@article{asareGitHubCopilotBad2023b,
  title = {Is {{GitHub}}’s {{Copilot}} as Bad as Humans at Introducing Vulnerabilities in Code?},
  author = {Asare, Owura and Nagappan, Meiyappan and Asokan, N.},
  date = {2023-09-23},
  journaltitle = {Empirical Software Engineering},
  shortjournal = {Empir Software Eng},
  volume = {28},
  number = {6},
  pages = {129},
  issn = {1573-7616},
  langid = {english}
}

@online{austinProgramSynthesisLarge2021b,
  title = {Program {{Synthesis}} with {{Large Language Models}}},
  author = {Austin, Jacob and Odena, Augustus and Nye, Maxwell and Bosma, Maarten and Michalewski, Henryk and Dohan, David and Jiang, Ellen and Cai, Carrie and Terry, Michael and Le, Quoc and Sutton, Charles},
  date = {2021-08-15},
  eprint = {2108.07732},
  eprinttype = {arxiv},
  eprintclass = {cs},
  pubstate = {preprint}
}

@online{baiConstitutionalAIHarmlessness2022b,
  title = {Constitutional {{AI}}: {{Harmlessness}} from {{AI Feedback}}},
  shorttitle = {Constitutional {{AI}}},
  author = {Bai, Yuntao and Kadavath, Saurav and Kundu, Sandipan and Askell, Amanda and Kernion, Jackson and Jones, Andy and Chen, Anna and Goldie, Anna and Mirhoseini, Azalia and McKinnon, Cameron and Chen, Carol and Olsson, Catherine and Olah, Christopher and Hernandez, Danny and Drain, Dawn and Ganguli, Deep and Li, Dustin and Tran-Johnson, Eli and Perez, Ethan and Kerr, Jamie and Mueller, Jared and Ladish, Jeffrey and Landau, Joshua and Ndousse, Kamal and Lukosuite, Kamile and Lovitt, Liane and Sellitto, Michael and Elhage, Nelson and Schiefer, Nicholas and Mercado, Noemi and DasSarma, Nova and Lasenby, Robert and Larson, Robin and Ringer, Sam and Johnston, Scott and Kravec, Shauna and Showk, Sheer El and Fort, Stanislav and Lanham, Tamera and Telleen-Lawton, Timothy and Conerly, Tom and Henighan, Tom and Hume, Tristan and Bowman, Samuel R. and Hatfield-Dodds, Zac and Mann, Ben and Amodei, Dario and Joseph, Nicholas and McCandlish, Sam and Brown, Tom and Kaplan, Jared},
  date = {2022-12-15},
  eprint = {2212.08073},
  eprinttype = {arxiv},
  eprintclass = {cs},
  pubstate = {preprint}
}

@article{barkeGroundedCopilotHow2023a,
  title = {Grounded {{Copilot}}: {{How Programmers Interact}} with {{Code-Generating Models}}},
  shorttitle = {Grounded {{Copilot}}},
  author = {Barke, Shraddha and James, Michael B. and Polikarpova, Nadia},
  date = {2023-04-06},
  journaltitle = {Proceedings of the ACM on Programming Languages},
  shortjournal = {Proc. ACM Program. Lang.},
  volume = {7},
  pages = {78:85--78:111},
  issue = {OOPSLA1}
}

@article{bommasaniHolisticEvaluationLanguage2023,
  title = {Holistic {{Evaluation}} of {{Language Models}}},
  author = {Bommasani, Rishi and Liang, Percy and Lee, Tony},
  date = {2023},
  journaltitle = {Annals of the New York Academy of Sciences},
  volume = {1525},
  number = {1},
  pages = {140--146},
  issn = {1749-6632},
  langid = {english}
}

@article{brierVerificationForecastsExpressed1950a,
  title = {Verification of Forecasts Expressed in Terms of Probability},
  author = {Brier, Glenn W.},
  date = {1950-01-01},
  journaltitle = {Monthly Weather Review},
  volume = {78},
  number = {1},
  pages = {1--3},
  publisher = {American Meteorological Society},
  issn = {1520-0493, 0027-0644},
  langid = {english}
}

@inproceedings{brownLanguageModelsAre2020b,
  title = {Language {{Models}} Are {{Few-Shot Learners}}},
  booktitle = {Advances in {{Neural Information Processing Systems}}},
  author = {Brown, Tom and Mann, Benjamin and Ryder, Nick and Subbiah, Melanie and Kaplan, Jared D and Dhariwal, Prafulla and Neelakantan, Arvind and Shyam, Pranav and Sastry, Girish and Askell, Amanda and Agarwal, Sandhini and Herbert-Voss, Ariel and Krueger, Gretchen and Henighan, Tom and Child, Rewon and Ramesh, Aditya and Ziegler, Daniel and Wu, Jeffrey and Winter, Clemens and Hesse, Chris and Chen, Mark and Sigler, Eric and Litwin, Mateusz and Gray, Scott and Chess, Benjamin and Clark, Jack and Berner, Christopher and McCandlish, Sam and Radford, Alec and Sutskever, Ilya and Amodei, Dario},
  date = {2020},
  volume = {33},
  pages = {1877--1901},
  publisher = {Curran Associates, Inc.}
}

@inproceedings{chenCloseLookCalibration2023b,
  title = {A {{Close Look}} into the {{Calibration}} of {{Pre-trained Language Models}}},
  booktitle = {Proceedings of the 61st {{Annual Meeting}} of the {{Association}} for {{Computational Linguistics}} ({{Volume}} 1: {{Long Papers}})},
  author = {Chen, Yangyi and Yuan, Lifan and Cui, Ganqu and Liu, Zhiyuan and Ji, Heng},
  editor = {Rogers, Anna and Boyd-Graber, Jordan and Okazaki, Naoaki},
  date = {2023-07},
  pages = {1343--1367},
  publisher = {Association for Computational Linguistics},
  location = {Toronto, Canada},
  booktitle = {{{ACL}} 2023}
}

@online{chenEvaluatingLargeLanguage2021b,
  title = {Evaluating {{Large Language Models Trained}} on {{Code}}},
  author = {Chen, Mark and Tworek, Jerry and Jun, Heewoo and Yuan, Qiming and Pinto, Henrique Ponde de Oliveira and Kaplan, Jared and Edwards, Harri and Burda, Yuri and Joseph, Nicholas and Brockman, Greg and Ray, Alex and Puri, Raul and Krueger, Gretchen and Petrov, Michael and Khlaaf, Heidy and Sastry, Girish and Mishkin, Pamela and Chan, Brooke and Gray, Scott and Ryder, Nick and Pavlov, Mikhail and Power, Alethea and Kaiser, Lukasz and Bavarian, Mohammad and Winter, Clemens and Tillet, Philippe and Such, Felipe Petroski and Cummings, Dave and Plappert, Matthias and Chantzis, Fotios and Barnes, Elizabeth and Herbert-Voss, Ariel and Guss, William Hebgen and Nichol, Alex and Paino, Alex and Tezak, Nikolas and Tang, Jie and Babuschkin, Igor and Balaji, Suchir and Jain, Shantanu and Saunders, William and Hesse, Christopher and Carr, Andrew N. and Leike, Jan and Achiam, Josh and Misra, Vedant and Morikawa, Evan and Radford, Alec and Knight, Matthew and Brundage, Miles and Murati, Mira and Mayer, Katie and Welinder, Peter and McGrew, Bob and Amodei, Dario and McCandlish, Sam and Sutskever, Ilya and Zaremba, Wojciech},
  date = {2021-07-14},
  eprint = {2107.03374},
  eprinttype = {arxiv},
  eprintclass = {cs},
  pubstate = {preprint}
}

@inproceedings{desaiCalibrationPretrainedTransformers2020b,
  title = {Calibration of {{Pre-trained Transformers}}},
  booktitle = {Proceedings of the 2020 {{Conference}} on {{Empirical Methods}} in {{Natural Language Processing}} ({{EMNLP}})},
  author = {Desai, Shrey and Durrett, Greg},
  editor = {Webber, Bonnie and Cohn, Trevor and He, Yulan and Liu, Yang},
  date = {2020-11},
  pages = {295--302},
  publisher = {Association for Computational Linguistics},
  location = {Online},
  eventtitle = {{{EMNLP}} 2020}
}

@inproceedings{fanAutomatedRepairPrograms2023b,
  title = {Automated {{Repair}} of {{Programs}} from {{Large Language Models}}},
  booktitle = {2023 {{IEEE}}/{{ACM}} 45th {{International Conference}} on {{Software Engineering}} ({{ICSE}})},
  author = {Fan, Zhiyu and Gao, Xiang and Mirchev, Martin and Roychoudhury, Abhik and Tan, Shin Hwei},
  date = {2023-05},
  pages = {1469--1481},
  issn = {1558-1225},
  eventtitle = {2023 {{IEEE}}/{{ACM}} 45th {{International Conference}} on {{Software Engineering}} ({{ICSE}})}
}

@article{gouesAutomatedProgramRepair2019a,
  title = {Automated Program Repair},
  author = {Goues, Claire Le and Pradel, Michael and Roychoudhury, Abhik},
  date = {2019-11-21},
  journaltitle = {Communications of the ACM},
  shortjournal = {Commun. ACM},
  volume = {62},
  number = {12},
  pages = {56--65},
  issn = {0001-0782}
}

@online{grosAISafetySubproblems2023b,
  title = {{{AI Safety Subproblems}} for {{Software Engineering Researchers}}},
  author = {Gros, David and Devanbu, Prem and Yu, Zhou},
  date = {2023-08-31},
  eprint = {2304.14597},
  eprinttype = {arxiv},
  eprintclass = {cs},
  pubstate = {preprint}
}

@inproceedings{guoCalibrationModernNeural2017b,
  title = {On {{Calibration}} of {{Modern Neural Networks}}},
  booktitle = {Proceedings of the 34th {{International Conference}} on {{Machine Learning}}},
  author = {Guo, Chuan and Pleiss, Geoff and Sun, Yu and Weinberger, Kilian Q.},
  date = {2017-07-17},
  pages = {1321--1330},
  publisher = {PMLR},
  issn = {2640-3498},
  eventtitle = {International {{Conference}} on {{Machine Learning}}},
  langid = {english}
}

@inproceedings{hendrycksUsingPreTrainingCan2019b,
  title = {Using {{Pre-Training Can Improve Model Robustness}} and {{Uncertainty}}},
  booktitle = {Proceedings of the 36th {{International Conference}} on {{Machine Learning}}},
  author = {Hendrycks, Dan and Lee, Kimin and Mazeika, Mantas},
  date = {2019-05-24},
  pages = {2712--2721},
  publisher = {PMLR},
  issn = {2640-3498},
  eventtitle = {International {{Conference}} on {{Machine Learning}}},
  eventtitle = {ICML},
  langid = {english}
}

@online{husainCodeSearchNetChallengeEvaluating2020a,
  title = {{{CodeSearchNet Challenge}}: {{Evaluating}} the {{State}} of {{Semantic Code Search}}},
  shorttitle = {{{CodeSearchNet Challenge}}},
  author = {Husain, Hamel and Wu, Ho-Hsiang and Gazit, Tiferet and Allamanis, Miltiadis and Brockschmidt, Marc},
  date = {2020-06-08},
  eprint = {1909.09436},
  eprinttype = {arxiv},
  pubstate = {preprint}
}

@inproceedings{islembouzeniaDyPyBenchBenchmarkExecutable2024,
  title = {{{DyPyBench}}: {{A}} Benchmark of Executable Python Software},
  booktitle = {Foundations of Software Engineering {{FSE}}},
  author = {Islem Bouzenia and Bajaj Piyush Krishan and Michael Pradel},
  date = {2024}
}

@inproceedings{izadiCodeFillMultitokenCode2022a,
  title = {{{CodeFill}}: Multi-Token Code Completion by Jointly Learning from Structure and Naming Sequences},
  shorttitle = {{{CodeFill}}},
  booktitle = {Proceedings of the 44th {{International Conference}} on {{Software Engineering}}},
  booktitle = {ICSE'22},
  author = {Izadi, Maliheh and Gismondi, Roberta and Gousios, Georgios},
  date = {2022-07-05},
  series = {{{ICSE}} '22},
  pages = {401--412},
  publisher = {Association for Computing Machinery},
  location = {New York, NY, USA},
  isbn = {978-1-4503-9221-1}
}

@inproceedings{jesseLargeLanguageModels2023b,
  title = {Large {{Language Models}} and {{Simple}}, {{Stupid Bugs}}},
  booktitle = {2023 {{IEEE}}/{{ACM}} 20th {{International Conference}} on {{Mining Software Repositories}} ({{MSR}})},
  author = {Jesse, Kevin and Ahmed, Toufique and Devanbu, Premkumar T. and Morgan, Emily},
  date = {2023-05},
  pages = {563--575},
  issn = {2574-3864},
  eventtitle = {2023 {{IEEE}}/{{ACM}} 20th {{International Conference}} on {{Mining Software Repositories}} ({{MSR}})}
}

@article{jiangHowCanWe2021a,
  title = {How {{Can We Know When Language Models Know}}? {{On}} the {{Calibration}} of {{Language Models}} for {{Question Answering}}},
  shorttitle = {How {{Can We Know When Language Models Know}}?},
  author = {Jiang, Zhengbao and Araki, Jun and Ding, Haibo and Neubig, Graham},
  date = {2021-09-08},
  journaltitle = {Transactions of the Association for Computational Linguistics},
  shortjournal = {Transactions of the Association for Computational Linguistics},
  volume = {9},
  pages = {962--977},
  issn = {2307-387X}
}

@inproceedings{jiangImpactCodeLanguage2023b,
  title = {Impact of {{Code Language Models}} on {{Automated Program Repair}}},
  booktitle = {2023 {{IEEE}}/{{ACM}} 45th {{International Conference}} on {{Software Engineering}} ({{ICSE}})},
  booktitle = {{{ICSE}}'23'},
  author = {Jiang, Nan and Liu, Kevin and Lutellier, Thibaud and Tan, Lin},
  date = {2023-05},
  pages = {1430--1442},
  issn = {1558-1225},
  eventtitle = {2023 {{IEEE}}/{{ACM}} 45th {{International Conference}} on {{Software Engineering}} ({{ICSE}})}
}

@inproceedings{justDefects4JDatabaseExisting2014a,
  title = {{{Defects4J}}: A Database of Existing Faults to Enable Controlled Testing Studies for {{Java}} Programs},
  shorttitle = {{{Defects4J}}},
  booktitle = {Proceedings of the 2014 {{International Symposium}} on {{Software Testing}} and {{Analysis}}},
  author = {Just, René and Jalali, Darioush and Ernst, Michael D.},
  date = {2014-07-21},
  booktitle = {{{ISSTA}} 2014},
  pages = {437--440},
  publisher = {Association for Computing Machinery},
  location = {New York, NY, USA},
  isbn = {978-1-4503-2645-2}
}

@online{kadavathLanguageModelsMostly2022c,
  title = {Language {{Models}} ({{Mostly}}) {{Know What They Know}}},
  author = {Kadavath, Saurav and Conerly, Tom and Askell, Amanda and Henighan, Tom and Drain, Dawn and Perez, Ethan and Schiefer, Nicholas and Hatfield-Dodds, Zac and DasSarma, Nova and Tran-Johnson, Eli and Johnston, Scott and El-Showk, Sheer and Jones, Andy and Elhage, Nelson and Hume, Tristan and Chen, Anna and Bai, Yuntao and Bowman, Sam and Fort, Stanislav and Ganguli, Deep and Hernandez, Danny and Jacobson, Josh and Kernion, Jackson and Kravec, Shauna and Lovitt, Liane and Ndousse, Kamal and Olsson, Catherine and Ringer, Sam and Amodei, Dario and Brown, Tom and Clark, Jack and Joseph, Nicholas and Mann, Ben and McCandlish, Sam and Olah, Chris and Kaplan, Jared},
  date = {2022-11-21},
  eprint = {2207.05221},
  eprinttype = {arxiv},
  eprintclass = {cs},
  pubstate = {preprint}
}

@inproceedings{karampatsisHowOftenSingleStatement2020a,
  title = {How {{Often Do Single-Statement Bugs Occur}}? {{The ManySStuBs4J Dataset}}},
  shorttitle = {How {{Often Do Single-Statement Bugs Occur}}?},
   booktitle = {Proceedings of the 17th {{International Conference}} on {{Mining Software Repositories}}},
  author = {Karampatsis, Rafael-Michael and Sutton, Charles},
  date = {2020-09-18},
  booktitle = {{{MSR}} '20},
  pages = {573--577},
  publisher = {Association for Computing Machinery},
  location = {New York, NY, USA},
  isbn = {978-1-4503-7517-7}
}

@online{keyTrustworthyNeuralProgram2023,
  title = {Toward {{Trustworthy Neural Program Synthesis}}},
  author = {Key, Darren and Li, Wen-Ding and Ellis, Kevin},
  date = {2023-10-09},
  eprint = {2210.00848},
  eprinttype = {arxiv},
  eprintclass = {cs},
  pubstate = {preprint}
}

@inproceedings{kimCodePredictionFeeding2021b,
  title = {Code {{Prediction}} by {{Feeding Trees}} to {{Transformers}}},
  booktitle = {2021 {{IEEE}}/{{ACM}} 43rd {{International Conference}} on {{Software Engineering}} ({{ICSE}})},
    booktitle = {ICSE'21},
  author = {Kim, Seohyun and Zhao, Jinman and Tian, Yuchi and Chandra, Satish},
  date = {2021-05},
  pages = {150--162},
  issn = {1558-1225},
  eventtitle = {2021 {{IEEE}}/{{ACM}} 43rd {{International Conference}} on {{Software Engineering}} ({{ICSE}})},
  eventtitle = {ICSE}
}

@inproceedings{kullTemperatureScalingObtaining2019,
  title = {Beyond Temperature Scaling: {{Obtaining}} Well-Calibrated Multi-Class Probabilities with {{Dirichlet}} Calibration},
  shorttitle = {Beyond Temperature Scaling},
  booktitle = {Advances in {{Neural Information Processing Systems}}},
  author = {Kull, Meelis and Perello Nieto, Miquel and Kängsepp, Markus and Silva Filho, Telmo and Song, Hao and Flach, Peter},
  date = {2019},
  volume = {32},
  publisher = {Curran Associates, Inc.}
}

@article{linTeachingModelsExpress2022b,
  title = {Teaching {{Models}} to {{Express Their Uncertainty}} in {{Words}}},
  author = {Lin, Stephanie and Hilton, Jacob and Evans, Owain},
  date = {2022-06-19},
  journaltitle = {Transactions on Machine Learning Research},
  issn = {2835-8856},
  langid = {english}
}

@inproceedings{liOperationalCalibrationDebugging2020a,
  title = {Operational Calibration: Debugging Confidence Errors for {{DNNs}} in the Field},
  shorttitle = {Operational Calibration},
  booktitle = {Proceedings of the 28th {{ACM Joint Meeting}} on {{European Software Engineering Conference}} and {{Symposium}} on the {{Foundations}} of {{Software Engineering}}},
  author = {Li, Zenan and Ma, Xiaoxing and Xu, Chang and Xu, Jingwei and Cao, Chun and Lü, Jian},
  date = {2020-11-08},
  series = {{{ESEC}}/{{FSE}} 2020},
  pages = {901--913},
  publisher = {Association for Computing Machinery},
  location = {New York, NY, USA},
  isbn = {978-1-4503-7043-1}
}

@article{liuYourCodeGenerated2023a,
  title = {Is {{Your Code Generated}} by {{ChatGPT Really Correct}}? {{Rigorous Evaluation}} of {{Large Language Models}} for {{Code Generation}}},
  shorttitle = {Is {{Your Code Generated}} by {{ChatGPT Really Correct}}?},
  author = {Liu, Jiawei and Xia, Chunqiu Steven and Wang, Yuyao and Zhang, Lingming},
  date = {2023-12-15},
  journaltitle = {Advances in Neural Information Processing Systems},
  volume = {36},
  pages = {21558--21572},

  langid = {english}
}

@online{loTrustworthySynergisticArtificial2023a,
  title = {Trustworthy and {{Synergistic Artificial Intelligence}} for {{Software Engineering}}: {{Vision}} and {{Roadmaps}}},
  shorttitle = {Trustworthy and {{Synergistic Artificial Intelligence}} for {{Software Engineering}}},
  author = {Lo, David},
  date = {2023-10-04},
  eprint = {2309.04142},
  eprinttype = {arxiv},
  eprintclass = {cs},
  pubstate = {preprint}
}

@article{luCodeXGLUEMachineLearning2021b,
  title = {{{CodeXGLUE}}: {{A Machine Learning Benchmark Dataset}} for {{Code Understanding}} and {{Generation}}},
  shorttitle = {{{CodeXGLUE}}},
  author = {Lu, Shuai and Guo, Daya and Ren, Shuo and Huang, Junjie and Svyatkovskiy, Alexey and Blanco, Ambrosio and Clement, Colin and Drain, Dawn and Jiang, Daxin and Tang, Duyu and Li, Ge and Zhou, Lidong and Shou, Linjun and Zhou, Long and Tufano, Michele and Gong, Ming and Zhou, Ming and Duan, Nan and Sundaresan, Neel and Deng, Shao Kun and Fu, Shengyu and Liu, Shujie},
  date = {2021-12-06},
  journaltitle = {NeurIPS},
  volume = {1},

  langid = {english}
}

@inproceedings{luoEmpiricalAnalysisFlaky2014a,
  title = {An Empirical Analysis of Flaky Tests},
  booktitle = {Proceedings of the 22nd {{ACM SIGSOFT International Symposium}} on {{Foundations}} of {{Software Engineering}}},
  author = {Luo, Qingzhou and Hariri, Farah and Eloussi, Lamyaa and Marinov, Darko},
  date = {2014-11-11},
  series = {{{FSE}} 2014},
  pages = {643--653},
  publisher = {Association for Computing Machinery},
  location = {New York, NY, USA},
  isbn = {978-1-4503-3056-5}
}

@inproceedings{mindererRevisitingCalibrationModern2021a,
  title = {Revisiting the {{Calibration}} of {{Modern Neural Networks}}},
  booktitle = {Advances in {{Neural Information Processing Systems}}},
  booktitle = {NeurIPS},
  author = {Minderer, Matthias and Djolonga, Josip and Romijnders, Rob and Hubis, Frances and Zhai, Xiaohua and Houlsby, Neil and Tran, Dustin and Lucic, Mario},
  date = {2021},
  volume = {34},
  pages = {15682--15694},
  publisher = {Curran Associates, Inc.}
}

@article{naeiniObtainingWellCalibrated2015a,
  title = {Obtaining {{Well Calibrated Probabilities Using Bayesian Binning}}},
  author = {Naeini, Mahdi Pakdaman and Cooper, Gregory and Hauskrecht, Milos},
  date = {2015-02-21},
  journaltitle = {Proceedings of the AAAI Conference on Artificial Intelligence},
  volume = {29},
  number = {1},
  issn = {2374-3468},
  issue = {1},
  langid = {english}
}

@article{niermanOutcomePredictionModel2001a,
  title = {Outcome Prediction Model for Very Elderly Critically Ill Patients},
  author = {Nierman, David M. and Schechter, Clyde B. and Cannon, Lisa M. and Meier, Diane E.},
  date = {2001-10},
  journaltitle = {Critical Care Medicine},
  volume = {29},
  number = {10},
  pages = {1853},
  issn = {0090=3493},
  langid = {american}
}

@online{nijkampCodeGen2LessonsTraining2023b,
  title = {{{CodeGen2}}: {{Lessons}} for {{Training LLMs}} on {{Programming}} and {{Natural Languages}}},
  shorttitle = {{{CodeGen2}}},
  author = {Nijkamp, Erik and Hayashi, Hiroaki and Xiong, Caiming and Savarese, Silvio and Zhou, Yingbo},
  date = {2023-07-11},
  eprint = {2305.02309},
  eprinttype = {arxiv},
  eprintclass = {cs},
  pubstate = {preprint}
}

@inproceedings{nixonMeasuringCalibrationDeep2019a,
  title = {Measuring {{Calibration}} in {{Deep Learning}}},
  author = {Nixon, Jeremy and Dusenberry, Michael W. and Zhang, Linchuan and Jerfel, Ghassen and Tran, Dustin},
  date = {2019},
  pages = {38--41},

  eventtitle = {Proceedings of the {{IEEE}}/{{CVF Conference}} on {{Computer Vision}} and {{Pattern Recognition Workshops}}}
}

@inproceedings{parkCalibrationPretrainedLanguage2022a,
  title = {On the {{Calibration}} of {{Pre-trained Language Models}} Using {{Mixup Guided}} by {{Area Under}} the {{Margin}} and {{Saliency}}},
  booktitle = {Proceedings of the 60th {{Annual Meeting}} of the {{Association}} for {{Computational Linguistics}} ({{Volume}} 1: {{Long Papers}})},
  author = {Park, Seo Yeon and Caragea, Cornelia},
  editor = {Muresan, Smaranda and Nakov, Preslav and Villavicencio, Aline},
  date = {2022-05},
  pages = {5364--5374},
  publisher = {Association for Computational Linguistics},
  location = {Dublin, Ireland},
  eventtitle = {{{ACL}} 2022}
}

@article{plattProbabilisticOutputsSupport1999b,
  title = {Probabilistic Outputs for Support Vector Machines and Comparisons to Regularized Likelihood Methods},
  author = {Platt, John and others},
  date = {1999},
  journaltitle = {Advances in large margin classifiers},
  volume = {10},
  number = {3},
  pages = {61--74},
  publisher = {Cambridge, MA}
}

@inproceedings{raychevProbabilisticModelCode2016b,
  title = {Probabilistic Model for Code with Decision Trees},
  booktitle = {Proceedings of the 2016 {{ACM SIGPLAN International Conference}} on {{Object-Oriented Programming}}, {{Systems}}, {{Languages}}, and {{Applications}}},
  author = {Raychev, Veselin and Bielik, Pavol and Vechev, Martin},
  date = {2016-10-19},
  booktitle = {{{OOPSLA}} 2016},
  pages = {731--747},
  publisher = {Association for Computing Machinery},
  location = {New York, NY, USA},
  isbn = {978-1-4503-4444-9}
}

@inproceedings{schusterYouAutocompleteMe2021b,
  title = {You {{Autocomplete Me}}: {{Poisoning Vulnerabilities}} in {{Neural Code Completion}}},
  shorttitle = {You {{Autocomplete Me}}},
  author = {Schuster, Roei and Song, Congzheng and Tromer, Eran and Shmatikov, Vitaly},
  date = {2021},
  pages = {1559--1575},

  eventtitle = {30th {{USENIX Security Symposium}} ({{USENIX Security}} 21)},
  isbn = {978-1-939133-24-3},
  langid = {english}
}

@article{schwarzGUESSProjectingMachine2019a,
  title = {{{GUESS}}: Projecting Machine Learning Scores to Well-Calibrated Probability Estimates for Clinical Decision-Making},
  shorttitle = {{{GUESS}}},
  author = {Schwarz, Johanna and Heider, Dominik},
  date = {2019-07-15},
  journaltitle = {Bioinformatics},
  shortjournal = {Bioinformatics},
  volume = {35},
  number = {14},
  pages = {2458--2465},
  issn = {1367-4803}
}

@online{shinnReflexionLanguageAgents2023b,
  title = {Reflexion: {{Language Agents}} with {{Verbal Reinforcement Learning}}},
  shorttitle = {Reflexion},
  author = {Shinn, Noah and Cassano, Federico and Berman, Edward and Gopinath, Ashwin and Narasimhan, Karthik and Yao, Shunyu},
  date = {2023-10-10},
  eprint = {2303.11366},
  eprinttype = {arxiv},
  eprintclass = {cs},
  pubstate = {preprint}
}

@online{srivastavaImitationGameQuantifying2023b,
  title = {Beyond the {{Imitation Game}}: {{Quantifying}} and Extrapolating the Capabilities of Language Models},
  shorttitle = {Beyond the {{Imitation Game}}},
  author = {Srivastava, Aarohi and Rastogi, Abhinav and Rao, Abhishek and Shoeb, Abu Awal Md and Abid, Abubakar and Fisch, Adam and Brown, Adam R. and Santoro, Adam and Gupta, Aditya and Garriga-Alonso, Adrià and Kluska, Agnieszka and Lewkowycz, Aitor and Agarwal, Akshat and Power, Alethea and Ray, Alex and Warstadt, Alex and Kocurek, Alexander W. and Safaya, Ali and Tazarv, Ali and Xiang, Alice and Parrish, Alicia and Nie, Allen and Hussain, Aman and Askell, Amanda and Dsouza, Amanda and Slone, Ambrose and Rahane, Ameet and Iyer, Anantharaman S. and Andreassen, Anders and Madotto, Andrea and Santilli, Andrea and Stuhlmüller, Andreas and Dai, Andrew and La, Andrew and Lampinen, Andrew and Zou, Andy and Jiang, Angela and Chen, Angelica and Vuong, Anh and Gupta, Animesh and Gottardi, Anna and Norelli, Antonio and Venkatesh, Anu and Gholamidavoodi, Arash and Tabassum, Arfa and Menezes, Arul and Kirubarajan, Arun and Mullokandov, Asher and Sabharwal, Ashish and Herrick, Austin and Efrat, Avia and Erdem, Aykut and Karakaş, Ayla and Roberts, B. Ryan and Loe, Bao Sheng and Zoph, Barret and Bojanowski, Bartłomiej and Özyurt, Batuhan and Hedayatnia, Behnam and Neyshabur, Behnam and Inden, Benjamin and Stein, Benno and Ekmekci, Berk and Lin, Bill Yuchen and Howald, Blake and Orinion, Bryan and Diao, Cameron and Dour, Cameron and Stinson, Catherine and Argueta, Cedrick and Ramírez, César Ferri and Singh, Chandan and Rathkopf, Charles and Meng, Chenlin and Baral, Chitta and Wu, Chiyu and Callison-Burch, Chris and Waites, Chris and Voigt, Christian and Manning, Christopher D. and Potts, Christopher and Ramirez, Cindy and Rivera, Clara E. and Siro, Clemencia and Raffel, Colin and Ashcraft, Courtney and Garbacea, Cristina and Sileo, Damien and Garrette, Dan and Hendrycks, Dan and Kilman, Dan and Roth, Dan and Freeman, Daniel and Khashabi, Daniel and Levy, Daniel and González, Daniel Moseguí and Perszyk, Danielle and Hernandez, Danny and Chen, Danqi and Ippolito, Daphne and Gilboa, Dar and Dohan, David and Drakard, David and Jurgens, David and Datta, Debajyoti and Ganguli, Deep and Emelin, Denis and Kleyko, Denis and Yuret, Deniz and Chen, Derek and Tam, Derek and Hupkes, Dieuwke and Misra, Diganta and Buzan, Dilyar and Mollo, Dimitri Coelho and Yang, Diyi and Lee, Dong-Ho and Schrader, Dylan and Shutova, Ekaterina and Cubuk, Ekin Dogus and Segal, Elad and Hagerman, Eleanor and Barnes, Elizabeth and Donoway, Elizabeth and Pavlick, Ellie and Rodola, Emanuele and Lam, Emma and Chu, Eric and Tang, Eric and Erdem, Erkut and Chang, Ernie and Chi, Ethan A. and Dyer, Ethan and Jerzak, Ethan and Kim, Ethan and Manyasi, Eunice Engefu and Zheltonozhskii, Evgenii and Xia, Fanyue and Siar, Fatemeh and Martínez-Plumed, Fernando and Happé, Francesca and Chollet, Francois and Rong, Frieda and Mishra, Gaurav and Winata, Genta Indra and family=Melo, given=Gerard, prefix=de, useprefix=true and Kruszewski, Germán and Parascandolo, Giambattista and Mariani, Giorgio and Wang, Gloria and Jaimovitch-López, Gonzalo and Betz, Gregor and Gur-Ari, Guy and Galijasevic, Hana and Kim, Hannah and Rashkin, Hannah and Hajishirzi, Hannaneh and Mehta, Harsh and Bogar, Hayden and Shevlin, Henry and Schütze, Hinrich and Yakura, Hiromu and Zhang, Hongming and Wong, Hugh Mee and Ng, Ian and Noble, Isaac and Jumelet, Jaap and Geissinger, Jack and Kernion, Jackson and Hilton, Jacob and Lee, Jaehoon and Fisac, Jaime Fernández and Simon, James B. and Koppel, James and Zheng, James and Zou, James and Kocoń, Jan and Thompson, Jana and Wingfield, Janelle and Kaplan, Jared and Radom, Jarema and Sohl-Dickstein, Jascha and Phang, Jason and Wei, Jason and Yosinski, Jason and Novikova, Jekaterina and Bosscher, Jelle and Marsh, Jennifer and Kim, Jeremy and Taal, Jeroen and Engel, Jesse and Alabi, Jesujoba and Xu, Jiacheng and Song, Jiaming and Tang, Jillian and Waweru, Joan and Burden, John and Miller, John and Balis, John U. and Batchelder, Jonathan and Berant, Jonathan and Frohberg, Jörg and Rozen, Jos and Hernandez-Orallo, Jose and Boudeman, Joseph and Guerr, Joseph and Jones, Joseph and Tenenbaum, Joshua B. and Rule, Joshua S. and Chua, Joyce and Kanclerz, Kamil and Livescu, Karen and Krauth, Karl and Gopalakrishnan, Karthik and Ignatyeva, Katerina and Markert, Katja and Dhole, Kaustubh D. and Gimpel, Kevin and Omondi, Kevin and Mathewson, Kory and Chiafullo, Kristen and Shkaruta, Ksenia and Shridhar, Kumar and McDonell, Kyle and Richardson, Kyle and Reynolds, Laria and Gao, Leo and Zhang, Li and Dugan, Liam and Qin, Lianhui and Contreras-Ochando, Lidia and Morency, Louis-Philippe and Moschella, Luca and Lam, Lucas and Noble, Lucy and Schmidt, Ludwig and He, Luheng and Colón, Luis Oliveros and Metz, Luke and Şenel, Lütfi Kerem and Bosma, Maarten and Sap, Maarten and family=Hoeve, given=Maartje, prefix=ter, useprefix=true and Farooqi, Maheen and Faruqui, Manaal and Mazeika, Mantas and Baturan, Marco and Marelli, Marco and Maru, Marco and Quintana, Maria Jose Ramírez and Tolkiehn, Marie and Giulianelli, Mario and Lewis, Martha and Potthast, Martin and Leavitt, Matthew L. and Hagen, Matthias and Schubert, Mátyás and Baitemirova, Medina Orduna and Arnaud, Melody and McElrath, Melvin and Yee, Michael A. and Cohen, Michael and Gu, Michael and Ivanitskiy, Michael and Starritt, Michael and Strube, Michael and Swędrowski, Michał and Bevilacqua, Michele and Yasunaga, Michihiro and Kale, Mihir and Cain, Mike and Xu, Mimee and Suzgun, Mirac and Walker, Mitch and Tiwari, Mo and Bansal, Mohit and Aminnaseri, Moin and Geva, Mor and Gheini, Mozhdeh and T, Mukund Varma and Peng, Nanyun and Chi, Nathan A. and Lee, Nayeon and Krakover, Neta Gur-Ari and Cameron, Nicholas and Roberts, Nicholas and Doiron, Nick and Martinez, Nicole and Nangia, Nikita and Deckers, Niklas and Muennighoff, Niklas and Keskar, Nitish Shirish and Iyer, Niveditha S. and Constant, Noah and Fiedel, Noah and Wen, Nuan and Zhang, Oliver and Agha, Omar and Elbaghdadi, Omar and Levy, Omer and Evans, Owain and Casares, Pablo Antonio Moreno and Doshi, Parth and Fung, Pascale and Liang, Paul Pu and Vicol, Paul and Alipoormolabashi, Pegah and Liao, Peiyuan and Liang, Percy and Chang, Peter and Eckersley, Peter and Htut, Phu Mon and Hwang, Pinyu and Miłkowski, Piotr and Patil, Piyush and Pezeshkpour, Pouya and Oli, Priti and Mei, Qiaozhu and Lyu, Qing and Chen, Qinlang and Banjade, Rabin and Rudolph, Rachel Etta and Gabriel, Raefer and Habacker, Rahel and Risco, Ramon and Millière, Raphaël and Garg, Rhythm and Barnes, Richard and Saurous, Rif A. and Arakawa, Riku and Raymaekers, Robbe and Frank, Robert and Sikand, Rohan and Novak, Roman and Sitelew, Roman and LeBras, Ronan and Liu, Rosanne and Jacobs, Rowan and Zhang, Rui and Salakhutdinov, Ruslan and Chi, Ryan and Lee, Ryan and Stovall, Ryan and Teehan, Ryan and Yang, Rylan and Singh, Sahib and Mohammad, Saif M. and Anand, Sajant and Dillavou, Sam and Shleifer, Sam and Wiseman, Sam and Gruetter, Samuel and Bowman, Samuel R. and Schoenholz, Samuel S. and Han, Sanghyun and Kwatra, Sanjeev and Rous, Sarah A. and Ghazarian, Sarik and Ghosh, Sayan and Casey, Sean and Bischoff, Sebastian and Gehrmann, Sebastian and Schuster, Sebastian and Sadeghi, Sepideh and Hamdan, Shadi and Zhou, Sharon and Srivastava, Shashank and Shi, Sherry and Singh, Shikhar and Asaadi, Shima and Gu, Shixiang Shane and Pachchigar, Shubh and Toshniwal, Shubham and Upadhyay, Shyam and Shyamolima and Debnath and Shakeri, Siamak and Thormeyer, Simon and Melzi, Simone and Reddy, Siva and Makini, Sneha Priscilla and Lee, Soo-Hwan and Torene, Spencer and Hatwar, Sriharsha and Dehaene, Stanislas and Divic, Stefan and Ermon, Stefano and Biderman, Stella and Lin, Stephanie and Prasad, Stephen and Piantadosi, Steven T. and Shieber, Stuart M. and Misherghi, Summer and Kiritchenko, Svetlana and Mishra, Swaroop and Linzen, Tal and Schuster, Tal and Li, Tao and Yu, Tao and Ali, Tariq and Hashimoto, Tatsu and Wu, Te-Lin and Desbordes, Théo and Rothschild, Theodore and Phan, Thomas and Wang, Tianle and Nkinyili, Tiberius and Schick, Timo and Kornev, Timofei and Tunduny, Titus and Gerstenberg, Tobias and Chang, Trenton and Neeraj, Trishala and Khot, Tushar and Shultz, Tyler and Shaham, Uri and Misra, Vedant and Demberg, Vera and Nyamai, Victoria and Raunak, Vikas and Ramasesh, Vinay and Prabhu, Vinay Uday and Padmakumar, Vishakh and Srikumar, Vivek and Fedus, William and Saunders, William and Zhang, William and Vossen, Wout and Ren, Xiang and Tong, Xiaoyu and Zhao, Xinran and Wu, Xinyi and Shen, Xudong and Yaghoobzadeh, Yadollah and Lakretz, Yair and Song, Yangqiu and Bahri, Yasaman and Choi, Yejin and Yang, Yichi and Hao, Yiding and Chen, Yifu and Belinkov, Yonatan and Hou, Yu and Hou, Yufang and Bai, Yuntao and Seid, Zachary and Zhao, Zhuoye and Wang, Zijian and Wang, Zijie J. and Wang, Zirui and Wu, Ziyi},
  date = {2023-06-12},
  eprint = {2206.04615},
  eprinttype = {arxiv},
  eprintclass = {cs, stat},
  pubstate = {preprint}
}

@article{steyerbergAssessingPerformancePrediction2010a,
  title = {Assessing the Performance of Prediction Models: A Framework for Some Traditional and Novel Measures},
  shorttitle = {Assessing the Performance of Prediction Models},
  author = {Steyerberg, Ewout W. and Vickers, Andrew J. and Cook, Nancy R. and Gerds, Thomas and Gonen, Mithat and Obuchowski, Nancy and Pencina, Michael J. and Kattan, Michael W.},
  date = {2010-01},
  journaltitle = {Epidemiology (Cambridge, Mass.)},
  shortjournal = {Epidemiology},
  volume = {21},
  number = {1},
  eprint = {20010215},
  eprinttype = {pmid},
  pages = {128--138},
  ssn = {1044-3983},
  pmcid = {PMC3575184}
}

@inproceedings{tianJustAskCalibration2023,
  %title = {Just {{Ask}} for {{Calibration}}: {{Strategies}} for {{Eliciting Calibrated Confidence Scores}} from {{Language Models Fine-Tuned}} with {{Human Feedback}}},
  title = {Just {{Ask}} for {{Calibration}}},
  booktitle = {Proceedings of the 2023 {{Conference}} on {{Empirical Methods}} in {{Natural Language Processing}}},
  author = {Tian, Katherine and Mitchell, Eric and Zhou, Allan and Sharma, Archit and Rafailov, Rafael and Yao, Huaxiu and Finn, Chelsea and Manning, Christopher},
  editor = {Bouamor, Houda and Pino, Juan and Bali, Kalika},
  date = {2023-12},
  pages = {5433--5442},
  publisher = {Association for Computational Linguistics},
  location = {Singapore},

  abstract = {A trustworthy real-world prediction system should produce well-calibrated confidence scores; that is, its confidence in an answer should be indicative of the likelihood that the answer is correct, enabling deferral to an expert in cases of low-confidence predictions. Recent studies have shown that unsupervised pre-training produces large language models (LMs) whose conditional probabilities are remarkably well-calibrated. However, the most widely-used LMs are fine-tuned with reinforcement learning from human feedback (RLHF-LMs), and some studies have suggested that RLHF-LMs produce conditional probabilities that are very poorly calibrated. In light of this perceived weakness, we conduct a broad evaluation of methods for extracting confidence scores from RLHF-LMs. For RLHF-LMs such as ChatGPT, GPT-4, and Claude, we find that verbalized confidences emitted as output tokens are typically better-calibrated than the model's conditional probabilities on the TriviaQA, SciQ, and TruthfulQA benchmarks, often reducing the expected calibration error by a relative 50\%.},
  eventtitle = {{{EMNLP}} 2023}
}

@article{weiChainofThoughtPromptingElicits2022,
  title = {Chain-of-{{Thought Prompting Elicits Reasoning}} in {{Large Language Models}}},
  author = {Wei, Jason and Wang, Xuezhi and Schuurmans, Dale and Bosma, Maarten and Ichter, Brian and Xia, Fei and Chi, Ed and Le, Quoc V. and Zhou, Denny},
  date = {2022-12-06},
  journaltitle = {Advances in Neural Information Processing Systems},
  volume = {35},
  pages = {24824--24837},

  langid = {english}
}

@inproceedings{zadroznyObtainingCalibratedProbability2001,
  title = {Obtaining Calibrated Probability Estimates from Decision Trees and Naive {{Bayesian}} Classifiers},
  booktitle = {Proceedings of the {{Eighteenth International Conference}} on {{Machine Learning}} ({{ICML}} 2001), {{Williams College}}, {{Williamstown}}, {{MA}}, {{USA}}, {{June}} 28 - {{July}} 1, 2001},
  booktitle={ICML'01},
  author = {Zadrozny, Bianca and Elkan, Charles},
  editor = {Brodley, Carla E. and Danyluk, Andrea Pohoreckyj},
  date = {2001},
  pages = {609--616},
  publisher = {Morgan Kaufmann}
}

@inproceedings{zadroznyTransformingClassifierScores2002,
  title = {Transforming Classifier Scores into Accurate Multiclass Probability Estimates},
  booktitle = {Proceedings of the Eighth {{ACM SIGKDD}} International Conference on {{Knowledge}} Discovery and Data Mining},
  author = {Zadrozny, Bianca and Elkan, Charles},
  date = {2002-07-23},
  booktitle = {{{KDD}} '02},
  pages = {694--699},
  publisher = {Association for Computing Machinery},
  location = {New York, NY, USA},
  isbn = {978-1-58113-567-1}
}

@inproceedings{zhangMixupEmpiricalRisk2018a,
  title = {Mixup: {{Beyond Empirical Risk Minimization}}},
  shorttitle = {Mixup},
  author = {Zhang, Hongyi and Cisse, Moustapha and Dauphin, Yann N. and Lopez-Paz, David},
  date = {2018-02-15},
  eventtitle = {International {{Conference}} on {{Learning Representations}}},
  langid = {english}
}

@online{zhangPACELMPromptingAugmentation2023b,
  title = {{{PACE-LM}}: {{Prompting}} and {{Augmentation}} for {{Calibrated Confidence Estimation}} with {{GPT-4}} in {{Cloud Incident Root Cause Analysis}}},
  shorttitle = {{{PACE-LM}}},
  author = {Zhang, Dylan and Zhang, Xuchao and Bansal, Chetan and Las-Casas, Pedro and Fonseca, Rodrigo and Rajmohan, Saravan},
  date = {2023-09-29},
  eprint = {2309.05833},
  eprinttype = {arxiv},
  eprintclass = {cs},
  pubstate = {preprint}
}

@article{zhangSurveyLearningbasedAutomated2023a,
  title = {A {{Survey}} of {{Learning-based Automated Program Repair}}},
  author = {Zhang, Quanjun and Fang, Chunrong and Ma, Yuxiang and Sun, Weisong and Chen, Zhenyu},
  date = {2023-12-23},
  journaltitle = {ACM Transactions on Software Engineering and Methodology},
  shortjournal = {ACM Trans. Softw. Eng. Methodol.},
  volume = {33},
  number = {2},
  pages = {55:1--55:69},
  issn = {1049-331X}
}

@online{zhengSurveyLargeLanguage2024,
  title = {A {{Survey}} of {{Large Language Models}} for {{Code}}: {{Evolution}}, {{Benchmarking}}, and {{Future Trends}}},
  shorttitle = {A {{Survey}} of {{Large Language Models}} for {{Code}}},
  author = {Zheng, Zibin and Ning, Kaiwen and Wang, Yanlin and Zhang, Jingwen and Zheng, Dewu and Ye, Mingxi and Chen, Jiachi},
  date = {2024-01-08},
  eprint = {2311.10372},
  eprinttype = {arxiv},
  eprintclass = {cs},
  pubstate = {preprint}
}

@inproceedings{zhouCalibrationPretrainedCode2024,
  title={On Calibration of Pre-trained Code Models},
  author={Zhou, Zhenhao and Sha, Chaofeng and Peng, Xin},
  booktitle={ICSE '24},
  booktitle={2024 IEEE/ACM 46th International Conference on Software Engineering (ICSE)},
  pages={861--861},
  year={2024},
  organization={IEEE Computer Society}
}

@inproceedings{zhouNavigatingGreyArea2023b,
  title = {Navigating the {{Grey Area}}: {{How Expressions}} of {{Uncertainty}} and {{Overconfidence Affect Language Models}}},
  shorttitle = {Navigating the {{Grey Area}}},
  booktitle = {Proceedings of the 2023 {{Conference}} on {{Empirical Methods}} in {{Natural Language Processing}}},
  booktitle = {EMNLP'23},
  author = {Zhou, Kaitlyn and Jurafsky, Dan and Hashimoto, Tatsunori},
  editor = {Bouamor, Houda and Pino, Juan and Bali, Kalika},
  date = {2023-12},
  pages = {5506--5524},
  publisher = {Association for Computational Linguistics},
  location = {Singapore},
  eventtitle = {{{EMNLP}} 2023}
}

@inproceedings{zieglerProductivityAssessmentNeural2022b,
  title = {Productivity Assessment of Neural Code Completion},
  booktitle = {Proceedings of the 6th {{ACM SIGPLAN International Symposium}} on {{Machine Programming}}},
  author = {Ziegler, Albert and Kalliamvakou, Eirini and Li, X. Alice and Rice, Andrew and Rifkin, Devon and Simister, Shawn and Sittampalam, Ganesh and Aftandilian, Edward},
  date = {2022-06-13},
  booktitle = {{{MAPS}} 2022},
  pages = {21--29},
  publisher = {Association for Computing Machinery},
  location = {New York, NY, USA},
  isbn = {978-1-4503-9273-0}
}

@INPROCEEDINGS{bm25fewshot,
  author={Nashid, Noor and Sintaha, Mifta and Mesbah, Ali},
  booktitle={2023 IEEE/ACM 45th International Conference on Software Engineering (ICSE)}, 
  title={Retrieval-Based Prompt Selection for Code-Related Few-Shot Learning}, 
  year={2023},
  volume={},
  number={},
  pages={2450-2462},
  doi={10.1109/ICSE48619.2023.00205}
}

@inproceedings{vasconcelos2022generation,
author = {Vasconcelos, Helena and Bansal, Gagan and Fourney, Adam and Liao, Q. Vera and Wortman Vaughan, Jennifer},
title = {Generation Probabilities are Not Enough: Improving Error Highlighting for AI Code Suggestions},
booktitle = {NeurIPS Workshop on Human-Centered AI},
year = {2022},
month = {10},
url = {https://www.microsoft.com/en-us/research/publication/generation-probabilities-are-not-enough-improving-error-highlighting-for-ai-code-suggestions/},
}

@inproceedings{rusure,
author = {Johnson, Daniel D. and Tarlow, Daniel and Walder, Christian},
title = {R-U-SURE? uncertainty-aware code suggestions by maximizing utility across random user intents},
year = {2023},
publisher = {JMLR.org},
booktitle = {Proceedings of the 40th International Conference on Machine Learning},
articleno = {623},
numpages = {45},
location = {Honolulu, Hawaii, USA},
series = {ICML'23}
}

@inproceedings{lever,
author = {Ni, Ansong and Iyer, Srini and Radev, Dragomir and Stoyanov, Ves and Yih, Wen-tau and Wang, Sida I. and Lin, Xi Victoria},
title = {LEVER: learning to verify language-to-code generation with execution},
year = {2023},
publisher = {JMLR.org},
booktitle = {Proceedings of the 40th International Conference on Machine Learning},
articleno = {1086},
numpages = {23},
location = {Honolulu, Hawaii, USA},
series = {ICML'23}
}

@online{lookleap,
      title={Look Before You Leap: An Exploratory Study of Uncertainty Measurement for Large Language Models}, 
      author={Yuheng Huang and Jiayang Song and Zhijie Wang and Shengming Zhao and Huaming Chen and Felix Juefei-Xu and Lei Ma},
      year={2023},
      eprint={2307.10236},
      archivePrefix={arXiv},
      primaryClass={cs.SE},
      %url={https://arxiv.org/abs/2307.10236}, 
}

@inproceedings{Liu2023LitCabLL,
  title={LitCab: Lightweight Language Model Calibration over Short- and Long-form Responses},
  author={Xin Liu and others},
  booktitle={ICML},
  year={2023},
  %url={https://api.semanticscholar.org/CorpusID:264811211}
}

@article{Liu2024EnhancingLM,
  title={Enhancing Language Model Factuality via Activation-Based Confidence Calibration and Guided Decoding},
  author={Xin Liu and others},
  journal={ArXiv},
  year={2024},
  volume={abs/2406.13230},
  %url={https://api.semanticscholar.org/CorpusID:270620078}
}

\clearpage
\crefalias{figure}{appendixfigure}
\setcounter{figure}{0}
\renewcommand{\thefigure}{A\arabic{figure}}

\setcounter{table}{0}
\renewcommand{\thetable}{A\arabic{table}}

\onecolumn
\appendix

\definecolor{improvement}{rgb}{1,0,0}
\definecolor{modelrow}{gray}{0.85}

\begin{table*}
    \centering
    \resizebox{0.6\textwidth}{!}{
        \begin{tblr}{
            colspec = {llccccccccc},
            vline{3,6,9} = {4-24}{gray!75, dotted},
            columns = {font=\small},
                }
            &  & \SetCell[c=3]{c} \textbf{Line Completion} & & & \SetCell[c=6]{c} \textbf{Program Repair} & & & & &  \\

            \cmidrule[gray!75]{3-6} \cmidrule[lr,gray!75]{6-11}
             & & \SetCell[c=3]{c} DyPyBench & & & \SetCell[c=3]{c} Defects4J & & & \SetCell[c=3]{c} SStubs \\
            \cmidrule[lr]{3-5} \cmidrule[lr]{6-8} \cmidrule[lr]{9-11}
            Model & Metric & ${\mathcal B}$ & $SS$ & $ECE$ & ${\mathcal B}$ & $SS$ & $ECE$ & ${\mathcal B}$ & $SS$ & $ECE$ \\
            \hline
GPT-3.5
    & Total Prob
        & \textbf{0.11} & \textbf{+0.38} & 0.03
        & 0.13 & +0.01 &
        & \textbf{0.16} & \textbf{+0.03} &
    \\
    & Avg Prob
        & 0.13 & +0.31 & 0.03
        & 0.13 & +0.01 &
        & 0.16 & +0.02 &
    \\
    & Ask T/F
        & 0.18 &  0.00 &
        & \textbf{0.12} & \textbf{+0.05} &
        & 0.16 &  0.00 &
    \\
    & Ask T/F N
        & 0.18 &  0.00 &
        & 0.13 & +0.02 &
        & 0.16 & +0.01 &
    \\
    & Verbalize
        & 0.18 &  0.00 &
        & 0.13 & -0.03 &
        & 0.16 &  0.00 &
    \\
    & Length
        & 0.17 & +0.06 & 0.03
        & 0.14 & -0.06 &
        & 0.16 &  0.00 &
    \\
    & Unskilled
        & 0.18 &  0.00 &
        & 0.13 &  0.00 &
        & 0.16 &  0.00 &
    \\
\hline[dashed]
Codex
    & Total Prob
        & \textbf{0.10} & \textbf{+0.43} & 0.02
        & \textbf{0.15} & \textbf{+0.01} &
        & \textbf{0.19} & \textbf{+0.05} & 0.02
    \\
    & Avg Prob
        & 0.12 & +0.34 & 0.02
        & 0.16 & -0.01 &
        & 0.19 & +0.05 & 0.02
    \\
    & Ask T/F
        & 0.18 & +0.01 &
        & 0.16 & -0.01 &
        & 0.19 & +0.04 &
    \\
    & Ask T/F N
        & 0.18 & +0.02 &
        & 0.16 & -0.01 &
        & 0.20 & +0.02 &
    \\
    & Verbalize
        & 0.18 &  0.00 &
        & 0.16 & -0.01 &
        & 0.20 &  0.00 &
    \\
    & Length
        & 0.16 & +0.09 & 0.02
        & 0.17 & -0.11 &
        & 0.20 &  0.00 &
    \\
    & Unskilled
        & 0.18 &  0.00 &
        & 0.16 &  0.00 &
        & 0.20 &  0.00 &
    \\
\hline[dashed]
CodeGen2
    & Total Prob
        & \textbf{0.09} & \textbf{+0.41} & 0.01
        & - & - &
        & - & - &
    \\
    & Avg Prob
        & 0.11 & +0.30 & 0.01
        & - & - &
        & - & - &
    \\
    & Ask T/F
        & 0.16 &  0.00 &
        & - & - &
        & - & - &
    \\
    & Ask T/F N
        & 0.16 &  0.00 &
        & - & - &
        & - & - &
    \\
    & Verbalize
        & 0.16 &  0.00 &
        & - & - &
        & - & - &
    \\
    & Length
        & 0.15 & +0.06 & 0.03
        & - & - &
        & - & - &
    \\
    & Unskilled
        & 0.16 &  0.00 &
        & - & - &
        & - & - &
    \\
\bottomrule
        \end{tblr}
    }
         \caption{Calibration measured as Platt-scaled Brier Score (${\mathcal B}$), Skill Score ($SS$), and Expected Calibration Error ($ECE$), with respect to ``exact-match'' (EM) notion of correctness, excluding function synthesis tasks as EM is not a useful or commonly used notion of correctness. In cases where the $SS$ is less than 0.05, the $ECE$ is omitted. This is because an estimate without any signal will become Platt-scaled to approximately the base rate. This will \emph{appear} as one well calibrated bin, resulting in an $ECE$ near zero, but does not provide information. CodeGen2 repair values are omitted as it does not perform the task with greater than 1\% accuracy.}
         \label{tab:brierExactMatchScaled}
     \vspace*{-2\baselineskip}
\end{table*}
\definecolor{improvement}{rgb}{1,0,0}
\definecolor{modelrow}{gray}{0.85}

\begin{table*}
    \centering
    \resizebox{0.6\textwidth}{!}{
        \begin{tblr}{
            colspec = {llccccc},
            vline{3,4,6} = {3-24}{gray!75, dotted},
            columns = {font=\small},
                }

            &  & \SetCell[c=1]{c} \textbf{Line Completion} & \SetCell[c=2]{c} \textbf{Function Synthesis} & &
            \SetCell[c=2]{c} \textbf{Program Repair} & \\
            \cmidrule[gray!75]{3} \cmidrule[lr,gray!75]{4-5} \cmidrule[l,gray!75]{6-8}

             Model & Metric & \SetCell[c=1]{c} DyPyBench &
             \SetCell[c=1]{c} HumanEval &
             \SetCell[c=1]{c} MBPP &
             \SetCell[c=1]{c} Defects4J &
             \SetCell[c=1]{c} SStubs \\
            \cmidrule[l]{1-17}
GPT-3.5
    & Total Prob
        & 0.67
        & 0.77
        & 0.70
        & 0.54
        & 0.61
    \\
    & Avg Prob
        & 0.68
        & 0.61
        & 0.56
        & 0.57
        & 0.60
    \\
    & Ask T/F
        & 0.54
        & 0.73
        & 0.71
        & 0.67
        & 0.54
    \\
    & Ask T/F N
        & 0.53
        & 0.74
        & 0.73
        & 0.67
        & 0.57
    \\
    & Verbalize
        & 0.53
        & 0.54
        & 0.60
        & 0.43
        & 0.51
    \\
    & Length
        & 0.53
        & 0.64
        & 0.61
        & 0.48
        & 0.53
    \\
    & Unskilled
        & 0.50
        & 0.50
        & 0.50
        & 0.50
        & 0.50
    \\
\hline[dashed]
Codex
    & Total Prob
        & 0.68
        & 0.70
        & 0.66
        & 0.52
        & 0.66
    \\
    & Avg Prob
        & 0.68
        & 0.71
        & 0.69
        & 0.57
        & 0.65
    \\
    & Ask T/F
        & 0.49
        & 0.65
        & 0.56
        & 0.62
        & 0.61
    \\
    & Ask T/F N
        & 0.49
        & 0.61
        & 0.53
        & 0.54
        & 0.59
    \\
    & Verbalize
        & 0.52
        & 0.46
        & 0.50
        & 0.45
        & 0.49
    \\
    & Length
        & 0.55
        & 0.51
        & 0.50
        & 0.43
        & 0.49
    \\
    & Unskilled
        & 0.50
        & 0.50
        & 0.50
        & 0.50
        & 0.50
    \\
\hline[dashed]
CodeGen2
    & Total Prob
        & 0.68
        & 0.44
        & 0.52
        & -
        & -
    \\
    & Avg Prob
        & 0.66
        & 0.67
        & 0.52
        & -
        & -
    \\
    & Ask T/F
        & 0.52
        & 0.39
        & 0.59
        & -
        & -
    \\
    & Ask T/F N
        & 0.54
        & 0.34
        & 0.57
        & -
        & -
    \\
    & Verbalize
        & 0.51
        & 0.49
        & 0.51
        & -
        & -
    \\
    & Length
        & 0.55
        & 0.39
        & 0.47
        & -
        & -
    \\
    & Unskilled
        & 0.50
        & 0.50
        & 0.50
        & -
        & -
    \\
\bottomrule
        \end{tblr}
    }
         \caption{\scriptsize AUC-ROC score of each technique}
         \label{tab:rocAllPassedRaw}
     \vspace*{-2\baselineskip}
\end{table*}

\begin{figure*}[h]
	\centering
	\includegraphics[width=0.7\textwidth, trim=0 2.5cm 0 0,scale=0.3]{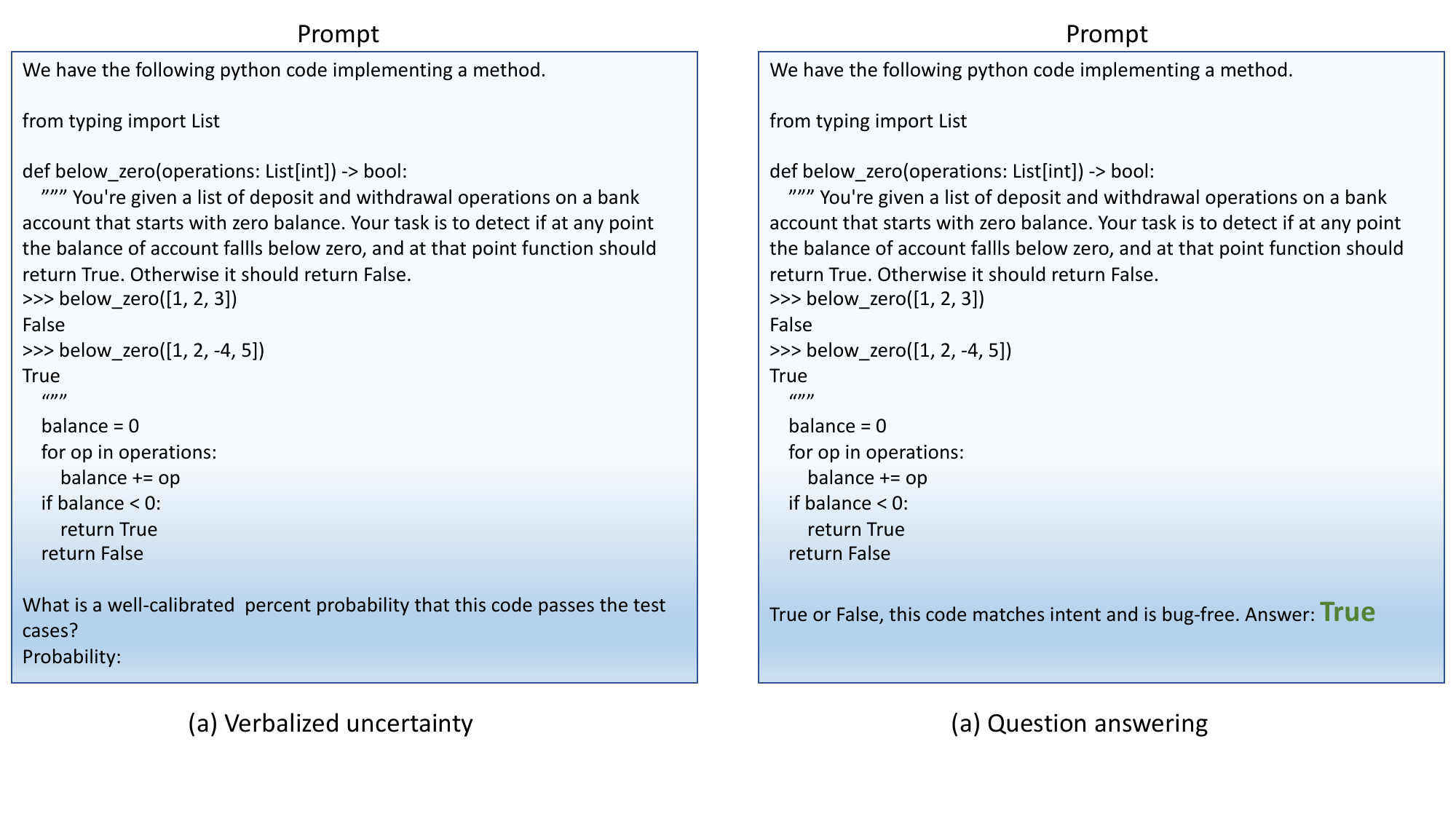}
	\caption{Prompts for Verbalized Self-Ask and Question Answering logit.}
	\label{app:selfask}
\end{figure*}

\begin{figure*}[ht]
	\centering
	\includegraphics[width=0.7\textwidth, trim=0 3cm 0 0,scale=0.40]{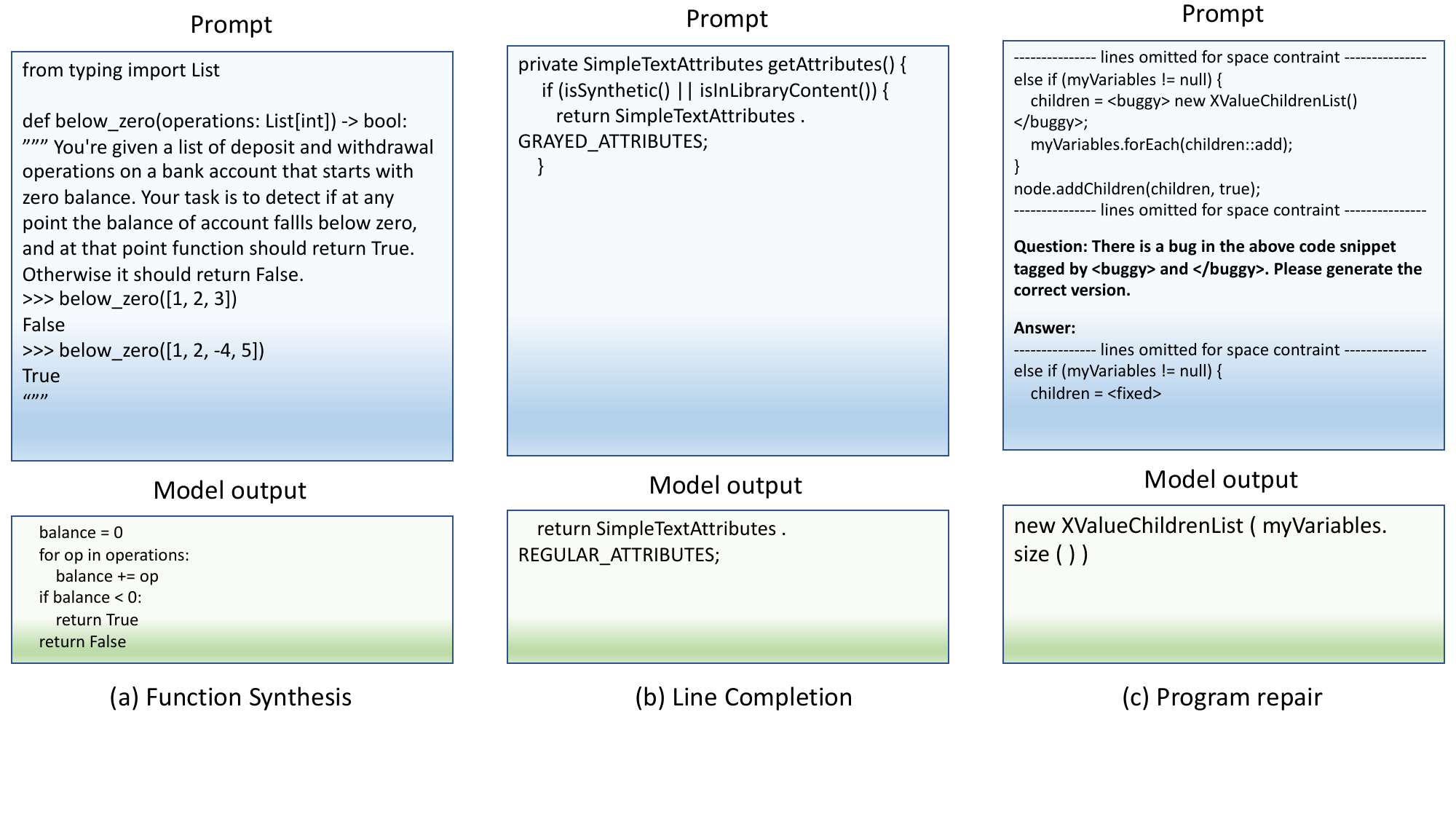}
	\caption{Prompt and model output for the tasks while calculating confidence measure based on Average Token Probability and Generated Sequence Probability.}
	\label{fig:logprob}
\end{figure*}

\begin{figure*}[ht]
	\centering
	\includegraphics[width=0.4\textwidth]{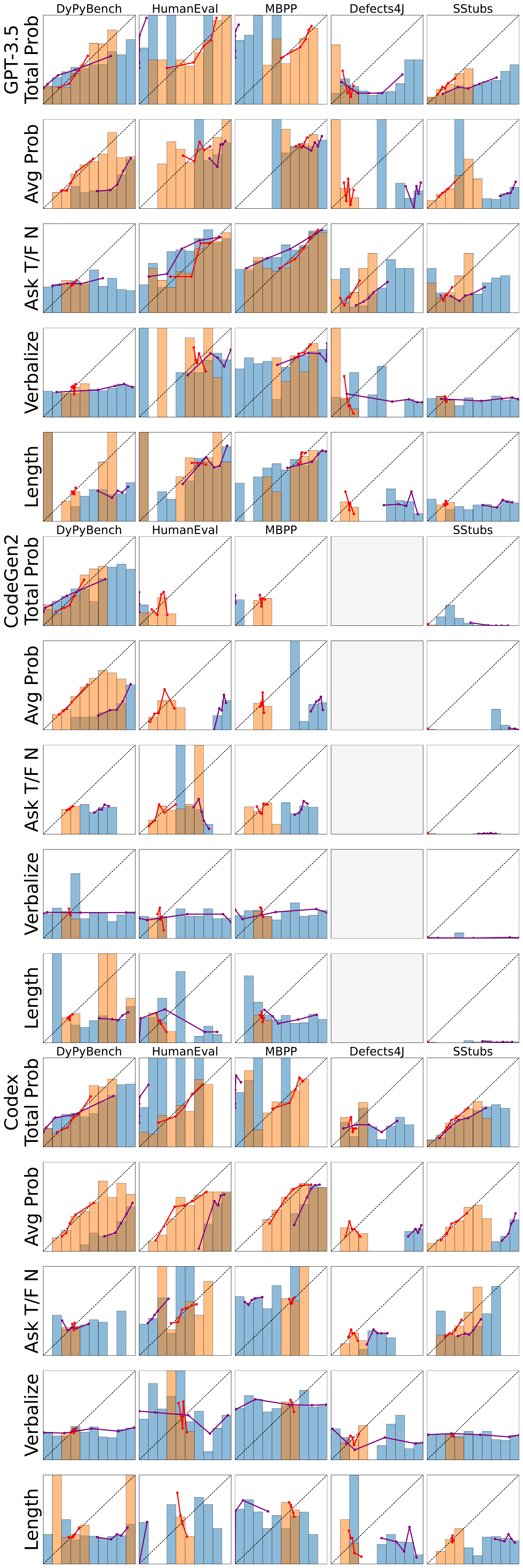}
	\caption{Calibration plots per model, confidence measure, and task. The blue bars represent nonscaled, evenly spaced bins. The orange bars are Platt scaled bins. The red lines represent five points of equal count quantiles (an equivalent number of problems in each bin).}
	\label{postageplot}
\end{figure*}

\begin{figure*}[ht]
	\centering
	\includegraphics[width=\textwidth]{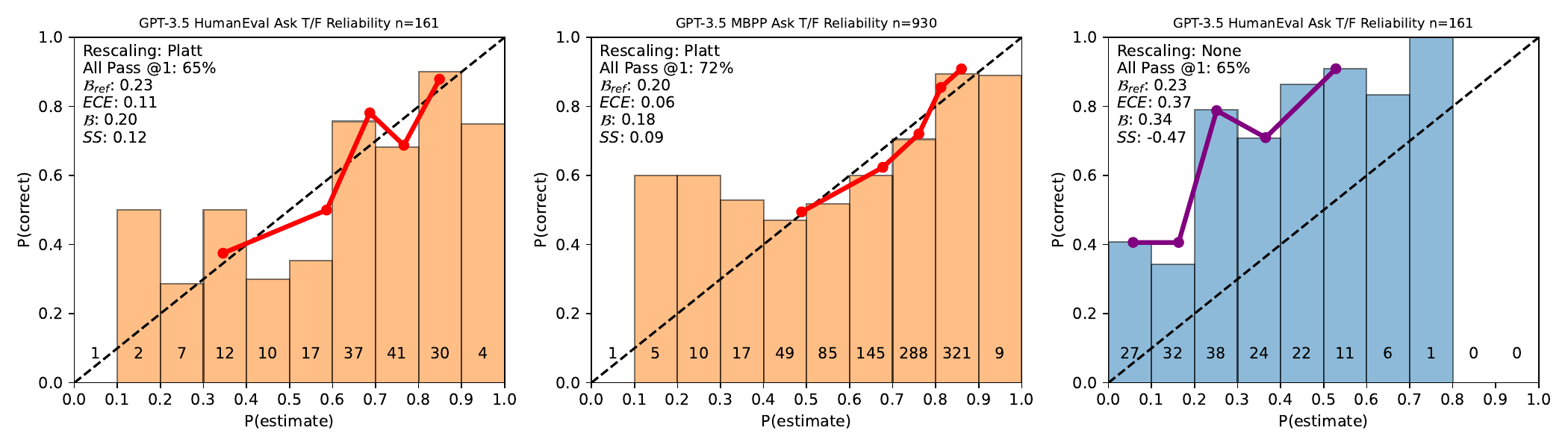}
	\caption{Reliability plots for GPT-3.5, from left to right: HumanEval Ask T/F (Scaled), MBPP Ask T/F (Scaled), and HumanEval Ask T/F (Nonscaled). Red line denotes five quantiles. All three examples have similar AUC (0.73, 0.71, 0.73) but vastly different $ECE$ (0.11, 0.06, 0.37).}
	\label{fig:differingAUC}
\end{figure*}

\begin{figure*}[ht]
	\centering
	\includegraphics[width=\textwidth]{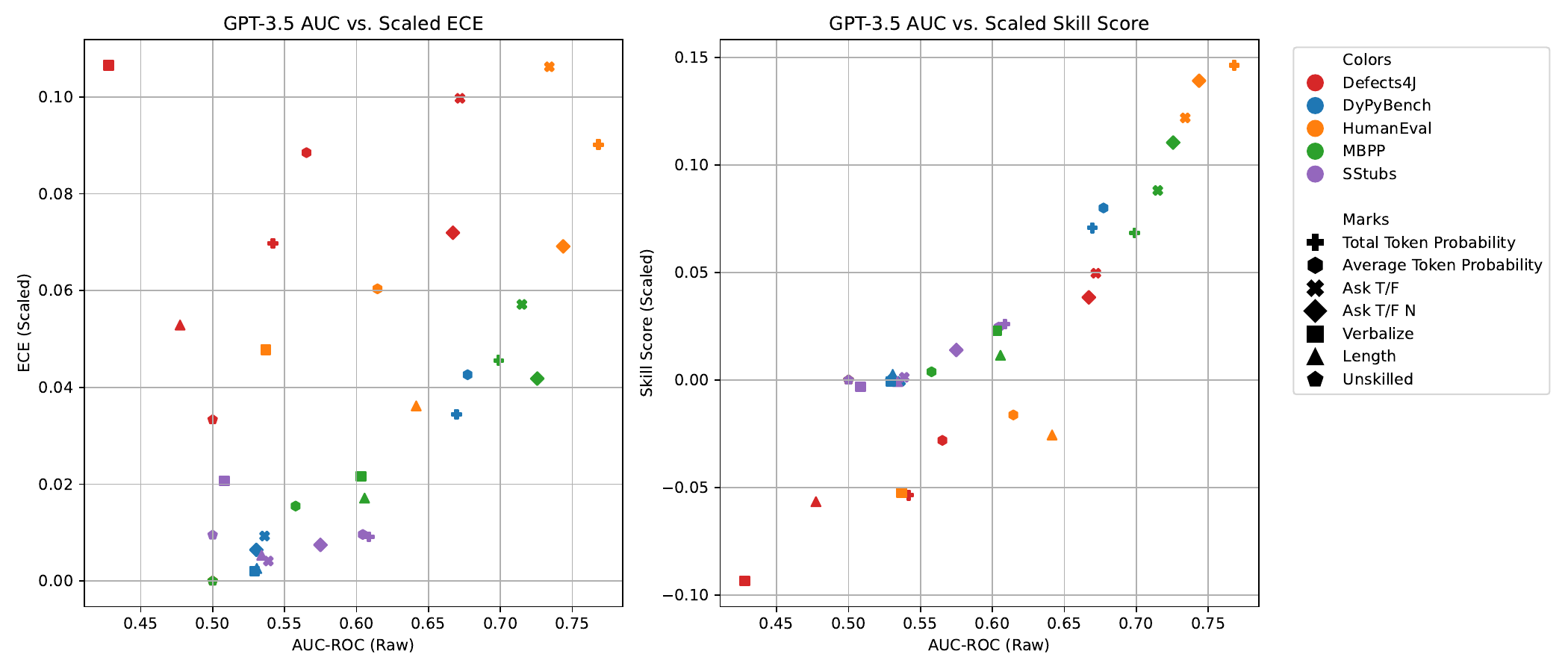}
	\caption{AUC vs Scaled $ECE$ (left) and AUC vs Scaled Skill Score (right) for GPT-3.5 confidence measures on all tasks. Shows a limited relationship between AUC and ECE, but a strong relationship between AUC and scaled SS.}
	\label{fig:AUCvsSSECE}
\end{figure*}

\begin{figure*}[ht]
	\centering
	\includegraphics[width=\textwidth]{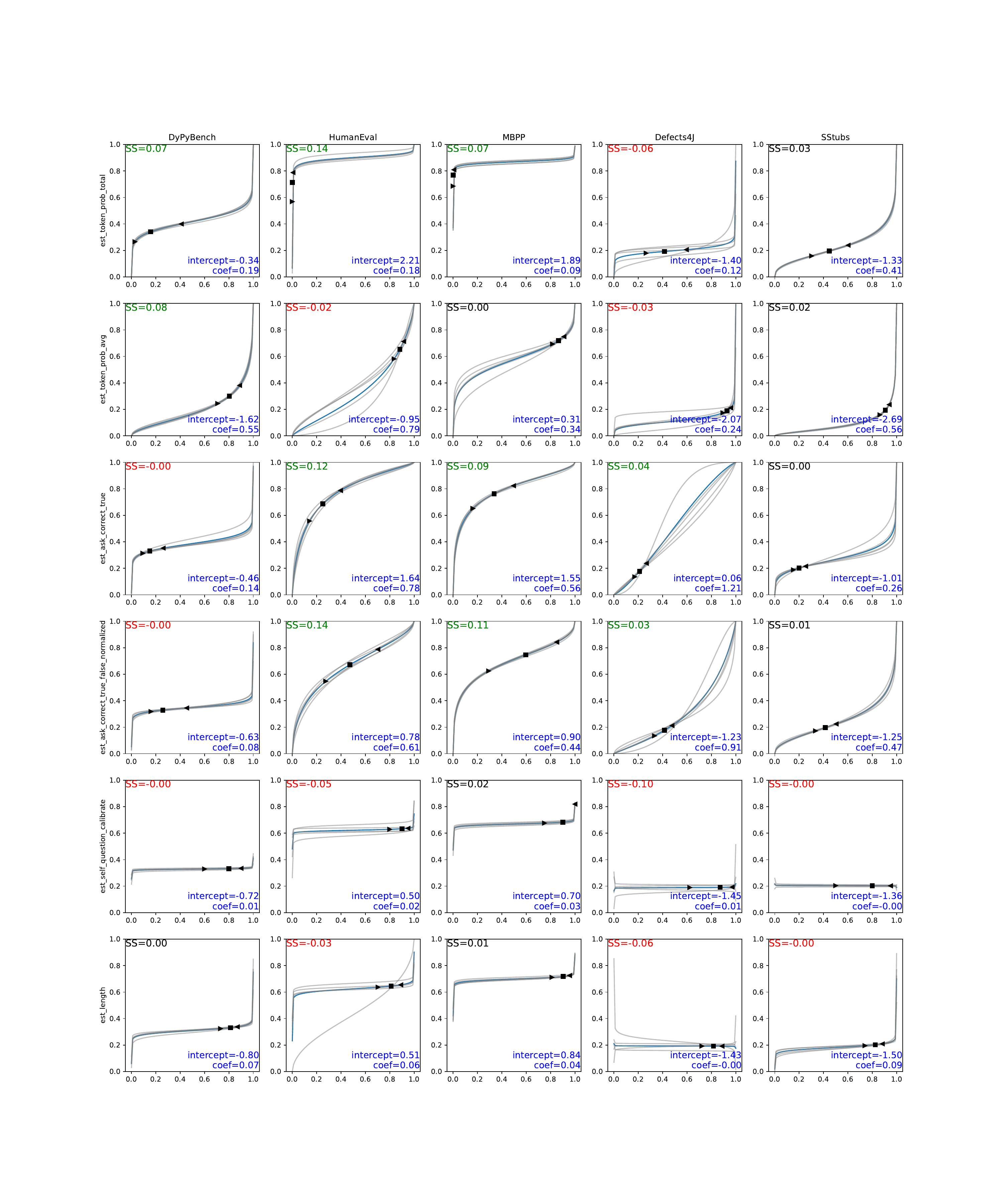}
	\caption{A comparison of the rescaling curves across tasks and measures for GPT-3.5. Logistic regression (Platt scaling) functions rescale the measurement (x-axis) to a new confidence (y-axis).
		The $\blacksquare$ represents a median measured value, $\blacktriangleright$ the lower quartile, and $\blacktriangleleft$ the upper quartile.
		The five curves from the different folds are shown in light gray, with the main line being the curve from fitting to all data. Dataset-measure pairs with less data or highly concentrated values have greater curve variance across folds. Scaled $SS$ is shown along with the logistic regression parameters.
	}
	\label{fig:rescalingFacet}
\end{figure*}

\definecolor{lightcoral}{rgb}{0.941, 0.502, 0.502}
\definecolor{greensmat}{rgb}{0.349, 0.702, 0.404}
\definecolor{purplesmat}{rgb}{0.4509, 0.43529, 0.67451}

\begin{figure*}[htbp]
	\centering

	\begin{subfigure}{0.49\textwidth}
		\includegraphics[width=\linewidth]{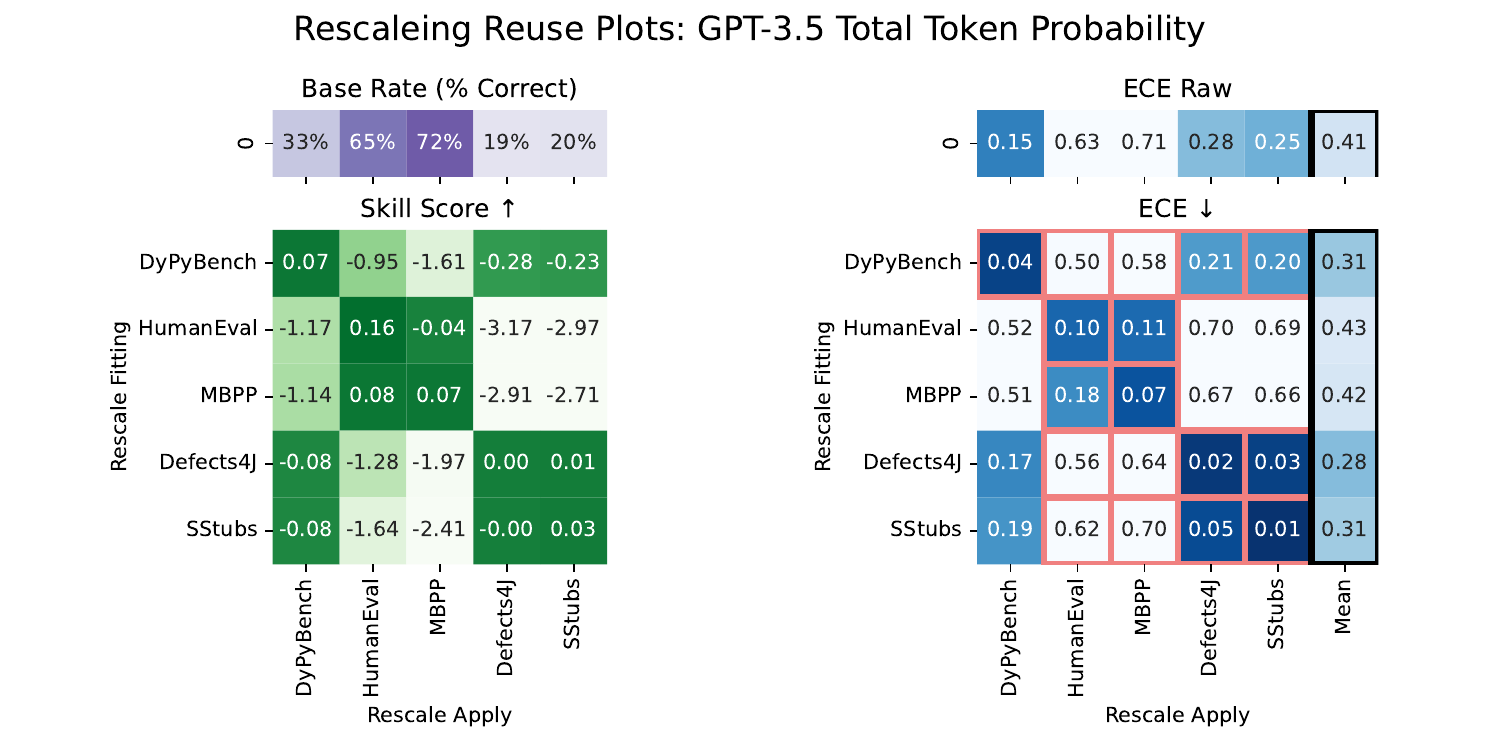}
		\label{fig:sub1}
	\end{subfigure}
	\begin{subfigure}{0.49\textwidth}
		\includegraphics[width=\linewidth]{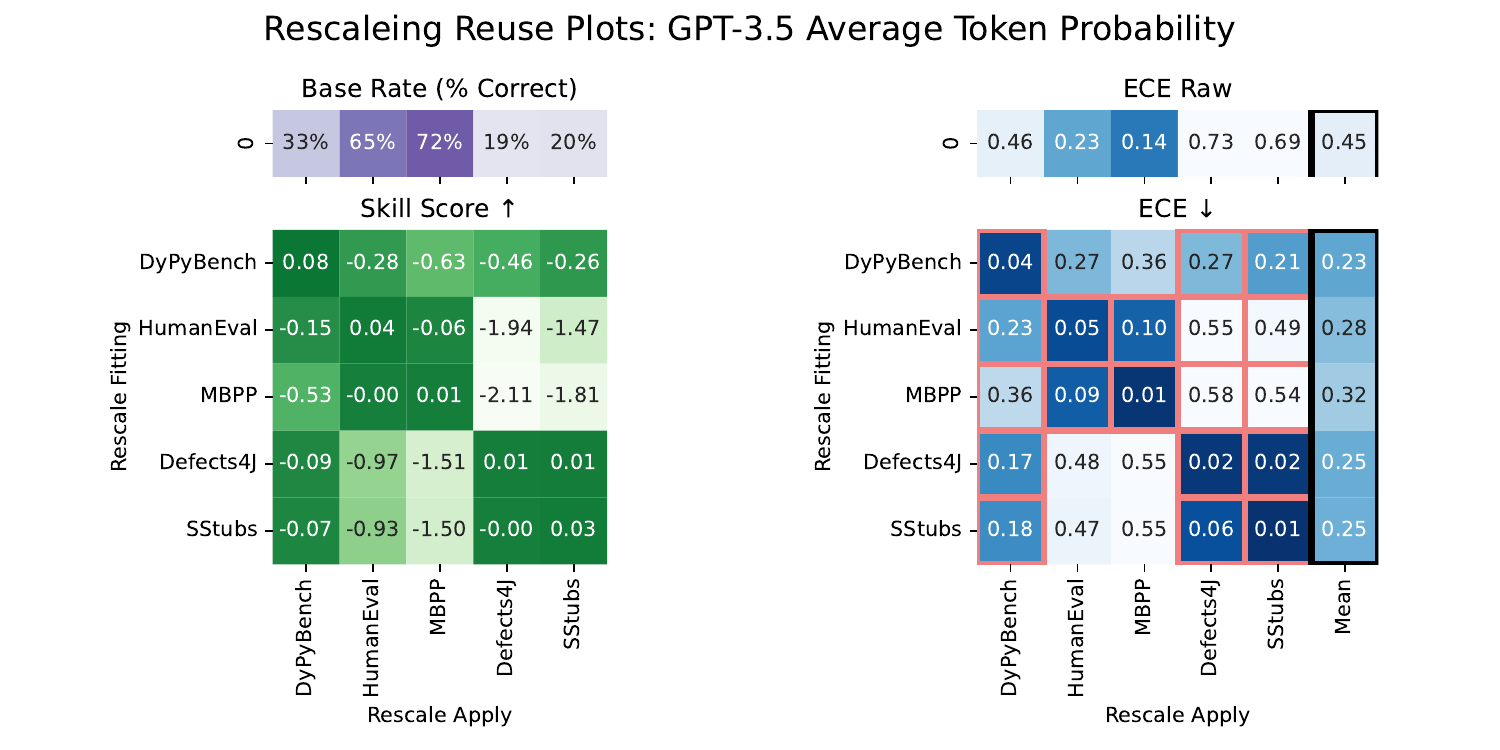}
		\label{fig:sub2}
	\end{subfigure}
	\begin{subfigure}{0.49\textwidth}
		\includegraphics[width=\linewidth]{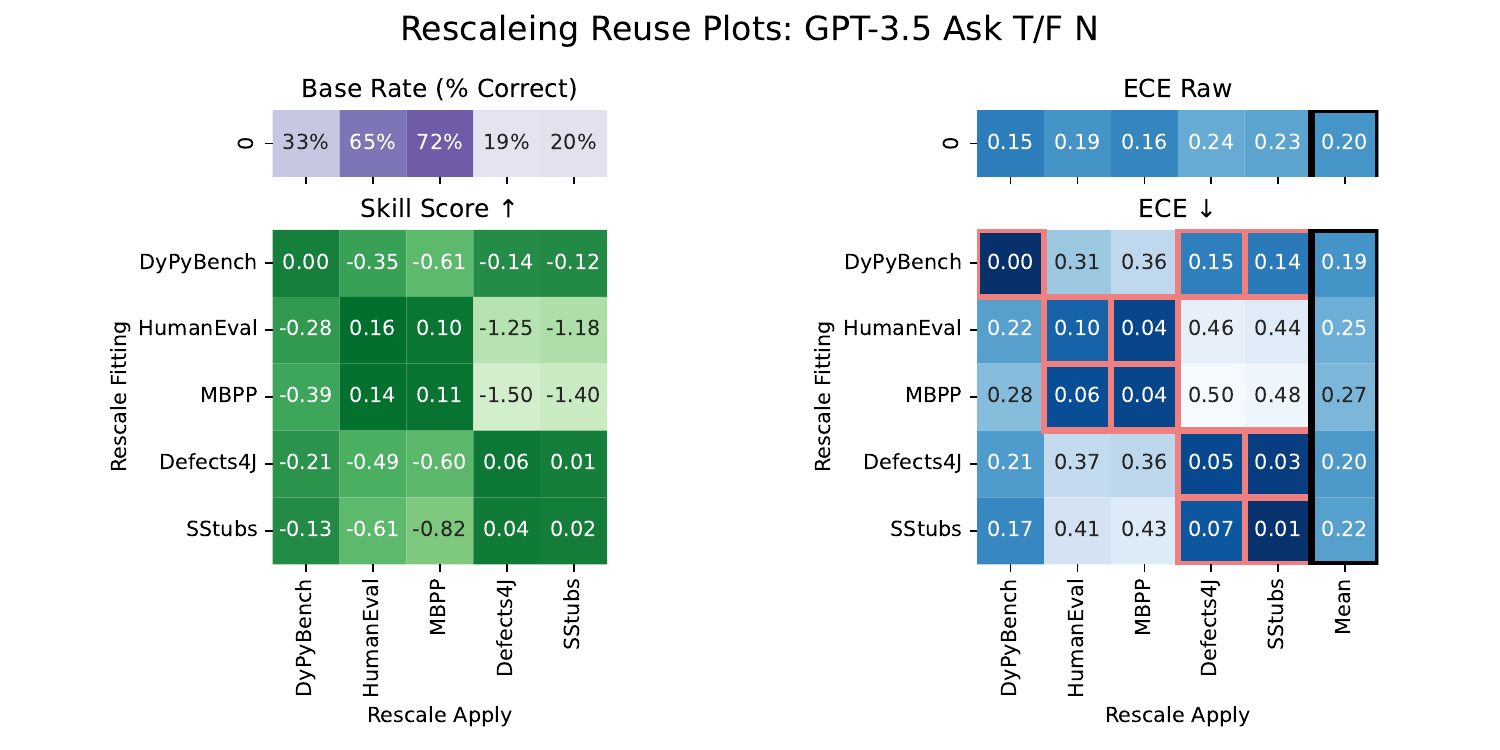}
		\label{fig:sub3}
	\end{subfigure}
	\begin{subfigure}{0.49\textwidth}
		\includegraphics[width=\linewidth]{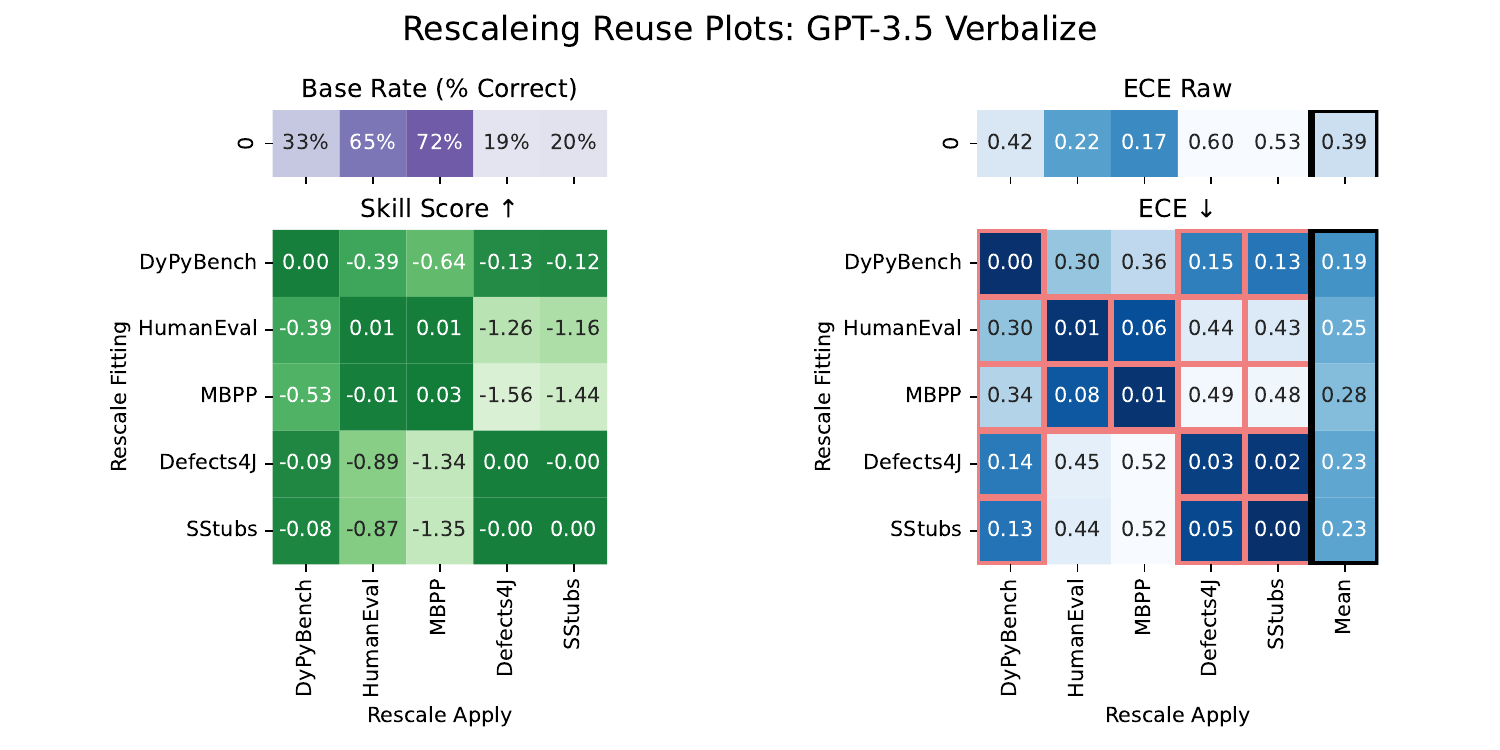}
		\label{fig:sub4}
	\end{subfigure}

	\caption{Exploration of GPT-3.5 confidences when fitting a rescaling for one the datasets, and then reusing it on another. In {\color{greensmat}green is Skill Score}, and in {\color{blue}blue is the $ECE$}. Above, we plot the raw (nonscaled) $ECE$ for each task. This informs whether a measure would be better calibrated if one uses it as-is, or one reuses the rescaling. Cells where there is an improvement in $ECE$ are shown in a {\color{lightcoral}coral} outline. Datasets within similar task \& {\color{purplesmat}base rate} exhibit most potential for reuse, but still liable to sizable changes in $SS$ or $ECE$. This analysis suggests that \emph{reflective} measures may be more robust across rescalings. Note values might differ slightly from results tables as the full data is used for training the rescaler, rather than folds.}
	\label{fig:reuseMatrix}
\end{figure*}

\clearpage

\begin{figure*}[!htb]
	\centering
	\includegraphics[width=\textwidth]{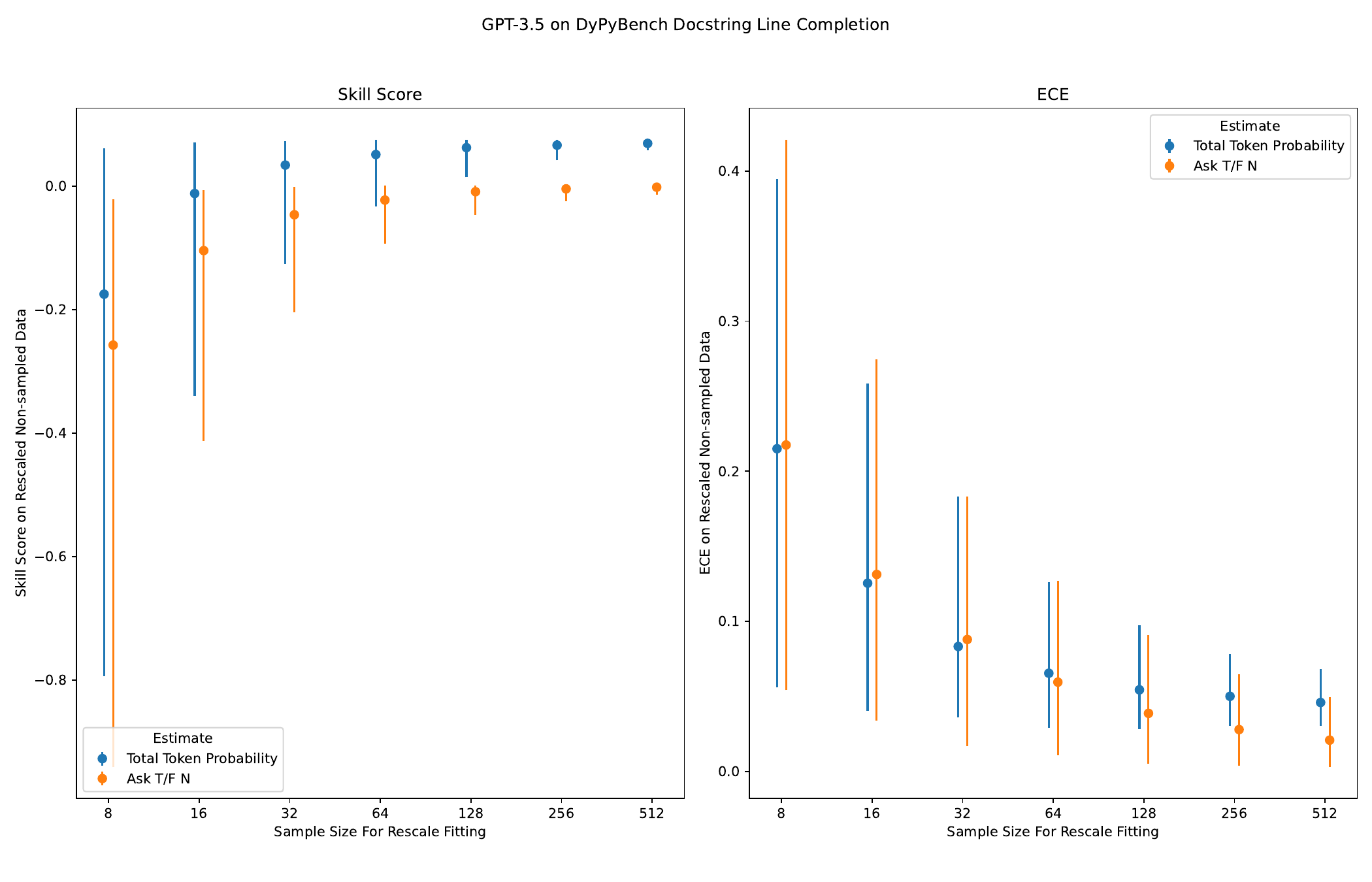}
	\caption{Bootstrapped resampling of varying sample for both an intrinsic and reflective measure. During each of 500 bootstrap simulations, a given number of data points is sampled. This is used to fit a Platt rescaling. We then apply that rescaling to the remaining non-sampled data points. We show the median simulation, and a 90\% interval. We observe that as the number of examples used for the rescaling increases, there are improvements in $SS$ and $ECE$.}
	\label{fig:bootstrapRescale}
\end{figure*}

\clearpage
\section*{Understanding effects of verbalized retry failures}\label{sec:verbalizeWacky}
Our implementation for verbalized confidence prompts the model to output the probability its generation is correct at temperature 1.0. If it does not contain a probability, then we resample up to 3 times. If that loop fails, then a confidence of 0.5 is returned.

This implementation seemed reasonable at the time (if the model won't tell you its confidence, just go with the maximum uncertainty of 50-50), but after collecting the data and analyzing results we reconsidered, as 50\% values might be overrepresented in the data.

To try to estimate how this might have influenced our results and conclusions, we searched for instances where the verbalized confidence was 50\% (this provides a upper bound on how often this happens. There can also be cases where the model actually verbalizes a 50\% confidence). This is relatively rare for GPT-3.5 (with the mean dataset having \GptMeanFracOfRowsFifty of instances as 50\%, range \GptMinFracOfRowsFifty-\GptMaxFracOfRowsFifty). It is more common for Codex (mean \CodexMeanFracOfRowsFifty, range \CodexMinFracOfRowsFifty-\CodexMaxFracOfRowsFifty) and for CodeGen2 (mean \CodegenMeanFracOfRowsFifty, range \CodegenMinFracOfRowsFifty-\CodegenMaxFracOfRowsFifty). This evidence of how the instruction tuned models are more likely to actually perform the prompted task.

We reran our analysis excluding all these instances. We do not believe a different handling of these fail-retry values would have greatly changed our conclusions. In the scaled case, the Skill Score on average did not change (mean diff of \MeanSkillScoreDiffScaled) with extreme change of \MinSkillScoreDiffScaled $SS$ when already low skill. In the nonscaled case there were some drops in calibration (mean $SS$ change of \MeanSkillScoreDiffNonscaled and mean $ECE$ change of \MeanEceDiffNonscaled). The more extreme changes areas of already poor calibration.

It is not clear what is the best default is in the situation where the model fails to verbalize a probability. It is not particularly valid to exclude these instance. More exploration is needed on this and the effects.

\vspace{1in}
\begin{table*}[!h]
	\centering
		\begin{tblr}{
				colspec = {lcccccc},
				columns = {font=\small},
			}

			                   & \SetCell[c=3]{c} HumanEval &               &                  & \SetCell[c=3]{c} MBPP                                       \\
			\cmidrule[gray!75]{2-5} \cmidrule[lr,gray!75]{5-8} \cmidrule[l,gray!75]{8-11}

			Confidence Measure & ${\mathcal B} \downarrow$  & $SS \uparrow$ & $ECE \downarrow$ & ${\mathcal B}\downarrow$ & $SS \uparrow$ & $ECE \downarrow$ \\
			\cmidrule[l]{1-11} 0-Shot Reflect
			                   & 0.23                       & 0.01          & 0.19
			                   & 0.22                       & -0.11         & 0.16                                                                           \\
			0-Shot Reflect (Scaled)
			                   & 0.20                       & 0.14          & 0.07
			                   & 0.18                       & 0.11          & 0.04                                                                           \\
			\cmidrule[l]{1-11}
			FS Random          & 0.20                       & 0.12          & 0.11             & 0.20                     & -0.03         & 0.15             \\
			FS Random (Scaled) & 0.19                       & 0.16          & 0.04             & 0.16                     & 0.16          & 0.04             \\
			FS BM25            & 0.19                       & 0.19          & 0.08             & 0.19                     & 0.00          & 0.11             \\
			FS BM25 (Scaled)   & 0.19                       & 0.18          & 0.04             & 0.17                     & 0.14          & 0.04             \\
			\bottomrule
		\end{tblr}
	\caption{Few-shot reflective prompting using \gptturbo. We observe the the unscaled skill score and ECE both improve. The raw SS improves 0.08-0.11 unscaled and further when scaled. The improvement from BM25 was more modest if doing rescaling, but appears useful if using raw values.}
	\label{table:fewshotfunc}
\end{table*}

\end{document}